\documentclass[10pt,journal,compsoc]{IEEEtran}
\usepackage{multirow}
\ifCLASSOPTIONcompsoc
  \usepackage[nocompress]{cite}
\else
  \usepackage{cite}
\fi
\ifCLASSINFOpdf
   \usepackage[pdftex]{graphicx}
\else
\fi
\usepackage{xcolor}

\usepackage{amsmath}
\usepackage{amssymb}
\usepackage{mathtools}
\usepackage{algorithmic}
\usepackage{array}
\newcolumntype{P}[1]{>{\centering\arraybackslash}m{#1}}

\definecolor{newcolor}{RGB}{59, 163, 0}
\newcommand{\new}[1]{#1}

\usepackage{url}
\begin{document}
\title{Integration-free Learning of Flow Maps}

\author{Saroj Sahoo and~Matthew Berger% <-this % stops a space
\IEEEcompsocitemizethanks{\IEEEcompsocthanksitem S. Sahoo and M. Berger are with the Department of Computer Science, Vanderbilt University.\protect\\
Contact email: matthew.berger@vanderbilt.edu}}
\IEEEtitleabstractindextext{%
\begin{abstract}

We present a method for learning neural representations of flow maps from time-varying vector field data. The flow map is pervasive within the area of flow visualization, as it is foundational to numerous visualization techniques, e.g. integral curve computation for pathlines or streaklines, as well as computing separation/attraction structures within the flow field. Yet bottlenecks in flow map computation, namely the numerical integration of vector fields, can easily inhibit their use within interactive visualization settings. In response, in our work we seek neural representations of flow maps that are efficient to evaluate, while remaining scalable to optimize, both in computation cost and data requirements. A key aspect of our approach is that we can frame the process of representation learning not in optimizing for samples of the flow map, but rather, a self-consistency criterion on flow map derivatives that eliminates the need for flow map samples, and thus numerical integration, altogether. Central to realizing this is a novel neural network design for flow maps, coupled with an optimization scheme, wherein our representation only requires the time-varying vector field for learning, encoded as instantaneous velocity. We show the benefits of our method over prior works in terms of accuracy and efficiency across a range of 2D and 3D time-varying vector fields, while showing how our neural representation of flow maps can benefit unsteady flow visualization techniques such as streaklines, and the finite-time Lyapunov exponent.
 
\end{abstract}

}

\maketitle

\IEEEdisplaynontitleabstractindextext

% \fi
\IEEEpeerreviewmaketitle

\IEEEraisesectionheading{\section{Introduction}\label{sec:introduction}}
Visual analysis is central to gaining insight on the underlying behaviour of unsteady flow data. Numerous visualization techniques have been developed to extract meaningful information from flow data, all in support of analyzing a variety of flow features. Among these techniques, notable ones include the finite-time Lyapunov exponents (FTLE)~\cite{haller2000finding}, used to understand the rate of separation between nearby particles integrated over a finite time interval, and its resulting Lagrangian coherent structures (LCS)~\cite{haller2000lagrangian, haller2000finding}, extracted as the ridges of the FTLE field. Streaklines are another visualization technique widely used by researchers, used to factor out background motion in flows, and identify underlying vortices that might be present. Other flow visualization techniques include line integral convolution (LIC)~\cite{cabral1993imaging}, almost invariant sets (AIS)~\cite{froyland2009almost}, finite-size Lyapunov exponent (FSLE)~\cite{artale1997dispersion, aurell1997predictability} and the coherent ergodic partitions~\cite{you2014eulerian}.

A core component common to all the above techniques is the computation of the flow map. The flow map provides the position of a particle advected under a flow over a finite time span, and typically, this is computed by integrating a time-varying vector field. For large time spans, this integration process can become computationally expensive, and thus impede interactivity within visual analysis. For example, trajectories of a dense set of particles sufficiently covering the spatial domain must be computed in order to generate the FTLE field. If the user is interested in interactively exploring FTLE under varying time spans, the expense in computing the flow map can hinder this exploration. Techniques that can improve the flow map computation time are thus attractive for a wide-variety of downstream visualization tasks.

In the literature, numerous techniques have been proposed for fast FTLE computation~\cite{garth2007efficient, kasten2009localized, sadlo2009visualizing, brunton2010fast, lipinski2010ridge, sadlo2011time}. Most of the these techniques fall into the general category of reducing the number of flow map evaluations required to accurately estimate the FTLE field. In a similar way, many other techniques have been proposed for fast LIC computation~\cite{stalling1995fast, battke1997fast}, and fast streakline computation~\cite{weinkauf2010streak}. However, most of these techniques are targeted towards improving the computation time of a specific downstream visualization task. For the majority of flow visualization techniques, the flow map computation time acts as a bottleneck, and few techniques have focused on the core of the problem i.e. improving the flow map computation time.

Motivated by these problems, in this work we propose a novel technique for fast and accurate flow map computation. We propose a novel coordinate-based neural network that serves as a surrogate for a flow map. Specifically, given a particle identified by a spatiotemporal coordinate, and time span, the network predicts the spatial position corresponding to the particle's integration under the flow field. Such neural representations of flow maps have been recently studied, both for 2D unsteady flows~\cite{han2021exploratory}, and more broadly for learning latent space representations~\cite{bilovs2021neural}. However, the learning of flow maps presents a number of challenges for existing methods. First, there is a steep training data requirement, where it is necessary to generate a large number of flow map samples on which to learn. Second, the input dimensionality varies over space, time, and time span, and thus in order to match the high dimensionality of the input space, the complexity (e.g. number of parameters) of the network often needs to be quite large. The complexity of the network can prove prohibitively expensive for training, provided the large dataset size, as well as prevent interactivity for use in downstream visual analysis.

Our approach aims to address, at once, these challenges through a novel network design that enables efficient inference, coupled with a novel optimization scheme for scalable training. A key aspect of our approach is that, through careful network design and optimization, we eliminate the need to learn from ground-truth flow map samples altogether. Rather, we take advantage of a basic property of flow maps: the instantaneous velocity of the flow map should be equivalent to the vector field. By optimizing the flow map derivative to represent the vector field, this ``primes'' the flow map itself to give a good approximation of particle transport, under small time spans. We show how to leverage this in devising a self-consistency criterion for learning the flow map under a range of time spans. Moreover, building on recent hybrid grid-MLP models~\cite{muller2022instant}, our network is efficient to evaluate, which enables both scalable training, as well as efficient inference. Our implementation can be found here : \url{https://github.com/SarojKumarSahoo/NIFM}

Our main contributions can be summarized as follows:
\begin{enumerate}
    \item We propose a novel network architecture for learning a neural representation of flow maps, that is fast, accurate and scalable.
    \item We propose a novel way to optimize the network only using the vector field, without requiring access to flow map samples during optimization.
    \item We show the advantage of using our technique by comparing against existing techniques both qualitatively and quantitatively. We demonstrate that, with modest training time, our method provides for a more accurate flow map approximation, and is more efficient at inference time, and hence applicable to numerous unsteady flow visualization techniques.
\end{enumerate}

\section{Related Work}
We introduce related work along four main directions: Lagrangian flow-based representations, methods for efficiently computing flow maps for FTLE, deep learning as applied to flow visualization, and more broadly relevant research in machine learning.

\textbf{Lagrangian Particle Interpolation.} The Lagrangian representation of unsteady flow fields stores data in the form of trajectories of massless particles. Agranovsky et al.~\cite{agranovsky2014improved} showed that, for exploratory analysis, in-situ trajectory computation and post-hoc interpolation is more storage-efficient than compared to traditional Eulerian representations. Several works to improve the accuracy of post-hoc Lagrangian particle interpolation have been proposed since~\cite{chandler2014interpolation, agranovsky2015multi, bujack2015lagrangian, sane2019interpolation, rapp2019void}. Bujack and Joy~\cite{bujack2015lagrangian} proposed a method for representing trajectories as parametric curves for a more accurate post-hoc interpolation, and additionally, they performed an error estimation of the proposed Lagrangian representation. In a similar way, several works focused on theoretical/empirical error analysis of Lagrangian interpolation~\cite{chandler2016analysis, hummel2016error, sane2018revisiting}. Even though these techniques do not require expensive numerical integration during post-hoc analysis, they are still expensive because of the number of steps required to compute the full trajectory. In the recent work by Li et al.~\cite{li2022efficient} they represented the trajectories as B-spline curves and improve the computation time of new trajectories by interpolation between the B-spline control points. Lagrangian representations of flow are attractive as a kind of data reduction, and assuming a sufficiently-dense sampling of trajectories, can often be quite accurate. Nevertheless, this representation can come at a steep computational cost for interactive analysis, as a common bottleneck is repeatedly performing spatial queries over irregularly-sampled particles in space-time.

%The goals of their work aligns with the goals of our work i.e. improving the computation time of flow maps, however, the type of data considered by both the techniques are different.

\textbf{Fast FTLE computation.} Computation of FTLE and its applications have received significant attention in the literature. Haller et al.~\cite{haller2000finding, haller2000lagrangian} showed that Lagrangian coherent structures can be extracted as the ridges of FTLE. Following this pioneering work many researchers focused on improving the computation time of FTLE. Garth et al.~\cite{garth2007efficient} proposed an incremental flow map approximation technique for improving the computation time of FTLE. Sadlo et al.~\cite{sadlo2007efficient} introduced a technique for FTLE ridge extraction using adaptive mesh refinement. Their proposed approach provides a speed-up in FTLE computation by avoiding integration of seed particles where no ridges are present. Kasten et al.~\cite{kasten2009localized} constructed a localized FTLE and additionally, a faster way to compute it by reusing the separation values from previous time steps. Lipinski et al.~\cite{lipinski2010ridge} proposed a ridge tracking algorithm that approximates ridges in the FTLE for each time step and then approximates the ridge location in subsequent times. Brunton et al.~\cite{brunton2010fast} proposed a fast FTLE computation technique taking advantage of flow map composition for longer flow map approximations. Hlawatsch et al.~\cite{hlawatsch2010hierarchical} introduced a hierarchical line integration scheme taking advantage of spatial and temporal coherence to improve the computation time of a dense set of particles. This work focuses on projection of particles based on the short pre-computed integral curves and thus has an accuracy trade-off. Sadlo et al.~\cite{sadlo2011time} proposed a grid advection technique for efficient FTLE computation taking advantage of temporal coherence. All these techniques are specifically targeted towards improving the computation time of a specific downstream task i.e. either FTLE or extraction of LCS from FTLE. In this work, we focus on improving the computation time of flow map itself allowing for a improved computation time for a wide range of downstream tasks.

\textbf{Deep learning for Flow Visualization.} In the recent years, numerous deep learning based techniques have been proposed in relation to flow visualization. Han et al.~\cite{han2019flow} introduced FlowNet an encoder-decoder deep learning framework for clustering, filtering and selection of streamlines and stream surfaces. Gu et al.~\cite{gu2021reconstructing} proposed a two-stage deep learning framework for flow field reconstruction using selected streamlines. Guo et. al~\cite{guo2020ssr} and Sahoo et al.~\cite{sahoo2021integration} proposed a deep learning based vector field super resolution using novel loss functions. Most relevant to our method is the flow map super resolution technique proposed by Jakob et al.~\cite{jakob2020fluid}, wherein the proposed technique requires a low resolution flow map as input and outputs higher resolution (4x) flow map. Even though the inference of the high resolution flow map is fast, the technique still the requires computation of a low resolution flow map making it less efficient in terms of overall computation time. Moreover, the flow maps are limited only to the grid locations. Our technique on the other hand is able to generate arbitrary flow map samples at any given space-time location. Another relevant work to our method is the recent work by Han et al.~\cite{han2021exploratory} where the authors proposed a deep learning technique which takes in a space-time coordinate and time span as input, and outputs the evaluation of the flow map. Our approach is similar to the technique proposed by Han et al. in scope, e.g. learning the full flow map of a corresponding vector field, however we differ in network architecture, optimization scheme, and data requirements. Specifically, our technique does not require flow map samples for supervision, and instead can learn a flow map representation only using a provided vector field.

\textbf{Neural Differential Equations} Neural ordinary differential equation (Neural ODE), a technique to solve initial-value ODE problems proposed by Chen et al.~\cite{chen2018neuralode} has been extended~\cite{liu2021second, norcliffe2021neural} and applied to various different research domains~\cite{han2021temporal, portwood2019turbulence}. Theoretically, since, flow maps are solution to an initial-value ODE, neural ODE should naturally extend to solve the problem. However, learning a flow map representation using neural ODE has not been studied yet. Our work is closely related to the work by Bilo{\v{s}} et al.~\cite{bilovs2021neural}, wherein they approximate the solution directly in a single step instead of integrating within a latent representation space. We draw inspiration from their work in the way we model the flow map prediction, however, the main distinction between our approach is in the network architecture design and the proposed optimization scheme. Our method of optimizing for flow map derivatives further draws inspiration from gradient-based learning methods, e.g. modeling shapes with gradient fields~\cite{cai2020learning}, and accelerating the volume-rendering integral through learning antiderivative networks (AutoInt) ~\cite{lindell2021autoint}. \new{A key difference between our method and AutoInt is that our novel network design allows us to forgo the requirement of computing integrals that are ultimately used for supervision in optimizing a gradient network, making our method of optimization more computationally efficient}.
\section{Methods}
In this section we present our approach, where we first describe the objective we seek to optimize, followed by a network design suited for this objective, and last we describe our specific approach to optimization.

\subsection{Integration-free learning}
\label{sec:integrationfree}
The flow map is an important mathematical tool that is utilized by numerous visualization techniques. To mathematically represent the flow map, let us consider a time-dependent flow field $\boldsymbol{\nu} : \mathbb{R}^n \times \mathbb{R} \rightarrow \mathbb{R}^n$, where $\boldsymbol{\nu}(\mathbf{x}(t), t)$ describes the vector of a particle at time $t$ with spatial position $\mathbf{x}(t) \in \mathbb{R}^n$. The trajectory of a mass-free particle, advected under the influence of the flow field $\boldsymbol{\nu}$, is governed by the following ordinary differential equation:
\begin{equation}
\frac{d \mathbf{x}(t)}{d t} = \boldsymbol{\nu}(\mathbf{x}(t),t) , \quad \mathbf{x}(t_0) = \mathbf{x}_0,
\end{equation}
where $\mathbf{x}_0$ represents the initial position of the particle at starting time $t_0$. Integrating this differential equation under a specified time span $\tau$ gives us the flow map $\Phi$, which varies in initial position $\mathbf{x}_0$, starting time $t_0$, and time span $\tau$:
\begin{equation}
    \Phi(\mathbf{x}_0, t_0, \tau): \mathbb{R}^n \times \mathbb{R} \times \mathbb{R} \rightarrow \mathbb{R}^n = \mathbf{x}_0 + \int_{t_0}^{t_0+\tau} \boldsymbol{\nu}(\mathbf{x}(t),t) dt.
\end{equation}
In practice, the computation of the flow map depends on (1) a means of interpolating the vector field at arbitrary space-time coordinates within the domain, and (2) a choice of numerical integration scheme, e.g. Euler or Runge-Kutta. Integrating for long time spans, however, can become a computational bottleneck when coupling the flow map with a particular visualization technique. This motivates the need for alternative flow map representations that can mitigate the expense of numerical integration.

In this work, we seek neural representations of flow maps, which we will denote $\hat{\Phi}$, that are (1) scalable to optimize, (2) efficient to evaluate, and (3) serve as accurate approximations. Satisfying all criteria, at once, is challenging with prior methods, as the choice of network design, objective(s) to be optimized, and data required for optimization, are all important considerations that interrelate. A standard approach~\cite{han2021exploratory,sahoo2022neural} is to collect a dense set of samples of the flow map, and optimize a neural network to reproduce these samples, either directly as its output~\cite{han2021exploratory}, or indirectly through integrating a learned vector field~\cite{sahoo2022neural}. However, to ensure good generalization, the number of flow map samples to collect needs to be quite large -- at least on the order of the vector field resolution -- with each sample requiring expensive numerical integration. Further, coordinate-based networks need to be sufficiently large for accurate learning, and thus combined, the dataset size and network complexity can lead to expensive training, and inefficient evaluation.

Our work foregoes the need for numerically integrating the vector field altogether. Instead, we optimize for \emph{flow map derivatives}, rather than the raw flow map output, taking advantage of the following basic property of a flow map:
\begin{equation}
    \label{eq:composition-exact}
    \frac{\partial \Phi(\mathbf{x}, t, \tau)}{\partial \tau} = \boldsymbol{\nu}(\Phi(\mathbf{x},t,\tau), t+\tau).
\end{equation}
In other words, the derivative of the flow map taken with respect to time span $\tau$, at position $\mathbf{x}$ and time $t$, can be found by (1) evaluating the flow map at the given inputs, and (2) accessing the vector field at the flow map's positional output, at time $t+\tau$. A flow map representation whose derivative is satisfied at all positions and times will, by construction, produce valid integral curves. Specifically, upon fixing position and time, evaluating the representation in increasing time span $\tau$ will yield a curve whose tangent vectors match the vector field as defined in Eq.~\eqref{eq:composition-exact}.

Of course, this approach assumes full access to the flow map itself, which is ultimately what we are trying to find. To help formulate a well-defined optimization problem, we identify two basic properties of a flow map that we expect any approximation should satisfy. Herein we refer to the neural flow map representation as $\hat{\Phi}$.

\begin{figure}[!t]
  \centering
 	\includegraphics[width=\linewidth]{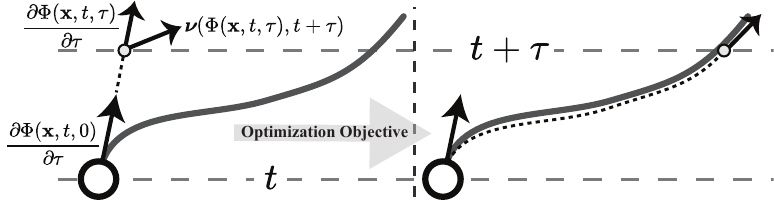}
	\caption{Our approach is based on a criterion of self-consistency, where the flow map derivative under some time span $\tau$ should match the flow map's instantaneous velocity at this location. A flow map whose instantaneous velocity initially matches the vector field can give us a linear approximation (left), violating self-consistency. By optimizing over a range of time spans, our approach aims to adjust the flow map such that this criterion is satisfied (right).}
    \label{fig:self-supervision}
\end{figure}

\textbf{(P1) Identity mapping.} When we integrate a particle for a time span of $\tau = 0$, then the flow map $\hat{\Phi}$ should return the starting position, irrespective of the starting time:
\begin{equation}
    \label{eq:identity}
    \hat{\Phi}(\mathbf{x}, t, 0) = \mathbf{x},
\end{equation}
We argue that any approximation $\hat{\Phi}$ should \emph{exactly} satisfy identity preservation. Otherwise, a small perturbation $\boldsymbol{\delta} \in \mathbb{R}^n$ yielding $\hat{\Phi}(\mathbf{x}, t, 0) = \mathbf{x}+\boldsymbol{\delta}$ would lead to an accumulation in error for repeated evaluation of the flow map approximation $\hat{\Phi}$.

\begin{figure*}[!t]
  \centering
 	\includegraphics[width=\linewidth]{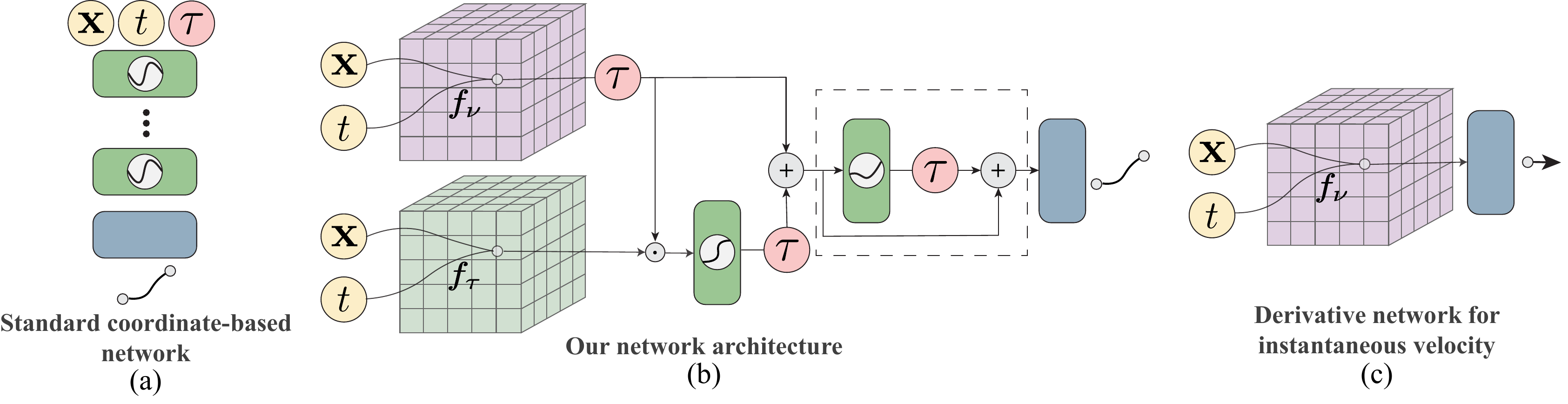}
	\caption{Standard coordinate-based networks (a) serve as simple models for flow maps, but are difficult to control and optimize. In contrast, our proposed network design (b) permits a clear delineation between the instantaneous velocity of the flow map, and integration for nonzero time spans ($\tau$), achieved via $\tau$-scaled residual connections. This leads to a simplified derivative network (c) at $\tau = 0$, one that is straightforward to fit to the vector field, and subsequently, stabilize flow map optimization.}
    \label{fig:illustrative_network}
\end{figure*}

\textbf{(P2) Instantaneous velocity.} For a time span of $\tau = 0$, if we compute the derivative of the flow map $\hat{\Phi}$ with respect to time span, then it should return the evaluation of the field $\boldsymbol{\nu}$ at the provided position $\mathbf{x}$ and time $t$:
\begin{equation}
    \label{eq:zero_tau_deriv}
    \frac{\partial \hat{\Phi}(\mathbf{x}, t, 0)}{\partial \tau} = \boldsymbol{\nu}(\mathbf{x}, t).
\end{equation}
A neural flow map representation $\hat{\Phi}$ whose derivative poorly approximates the vector field, e.g. points in a different direction, can lead to particle trajectories that diverge from actual trajectories. Indeed, upon a simple first-order approximation, we have:
\begin{equation}
    \label{eq:zero_advect}
    \hat{\Phi}(\mathbf{x}, t, \varepsilon) \approx \hat{\Phi}(\mathbf{x}, t, 0) + \varepsilon \boldsymbol{\nu}(\mathbf{x}, t).
\end{equation}
and thus, if the derivative of $\hat{\Phi}$ at $\tau = 0$ is poorly approximated, then this negatively impacts the action of the flow map itself. Also note that failing to preserve the identity mapping \textbf{(P1)} can further compound error.

Assuming the above properties hold, we propose the following criterion of \emph{self-consistency} for learning flow maps:
\begin{equation}
    \label{eq:composition-detail}
    l_s(\mathbf{x},t,\tau) = \bigg\lVert \frac{\partial \hat{\Phi}(\mathbf{x}, t, \tau)}{\partial \tau} - \frac{\partial \hat{\Phi}(\hat{\Phi}(\mathbf{x}, t, \tau),t+\tau,0)}{\partial \tau} \bigg\rVert.
\end{equation}
Here we have replaced the vector field in Eq.~\eqref{eq:composition-exact} with the flow map derivative. Hence, assuming property \textbf{(P2)} holds, the derivative of the flow map at $\tau = 0$ will faithfully represent the vector field. By minimizing this objective over the full domain via:
\begin{equation}
    \label{eq:composition-expectation}
    \mathcal{L}_s = \mathbb{E}_{(\mathbf{x},t) \in \mathcal{D}, \tau \in \mathcal{T}} \left[ l_s(\mathbf{x},t,\tau) \right],
\end{equation}
where $\mathcal{D}$ is the spatiotemporal domain, and $\mathcal{T} = [\tau_{min}, \tau_{max}]$ is an interval of time spans we aim to support in our approximation, we can ensure global self-consistency. Such a property is fundamental to \emph{any} flow map, but it is possible for $\hat{\Phi}$ to minimize Eq.~\eqref{eq:composition-expectation}, while remaining a poor approximation of $\Phi$. However, if instantaneous velocity is well-satisfied \textbf{(P2)}, and remains fixed, if not minimally changed, during optimization, then this will limit the space of flow maps that satisfy Eq.~\eqref{eq:composition-expectation}.

To provide intuition for our approach, if a flow map initially satisfies properties \textbf{(P1)} and \textbf{(P2)} then this can give a simple linear approximation, as shown in Fig.~\ref{fig:self-supervision} (left). However, the self-consistency criterion will naturally report a high loss for a sufficiently-large $\tau > 0$. By optimizing over a range of time spans $\mathcal{T}$, we can incrementally improve on self-consistency: first for small time spans, given \textbf{(P2)} holds, and then for larger time spans, as notionally depicted in Fig.~\ref{fig:self-supervision} (right). This idea of incrementally building the flow map has precedence in the literature~\cite{hlawatsch2010hierarchical}, but in our approach we eliminate the need for numerical integration, and instead \emph{only} require access to the original vector field. But critical to our approach, we require that the flow map approximation satisfy properties \textbf{(P1)} and \textbf{(P2)}. We next turn to a novel network design suited for these ends.

\subsection{A network design for flow maps}
\label{sec:networkdesign}

Coordinate-based neural networks, in particular ones based on sinusoidal positional encodings~\cite{sitzmann2020implicit,tancik2020fourier,fathony2021multiplicative}, are a natural choice for our flow map network design. Specifically, position ($n$ dimensions), time (1 dimension), and time span (1 dimension) can collectively be treated as individual coordinates as input to a multi-layer perceptron (MLP)~\cite{sitzmann2020implicit,tancik2020fourier}, whose output corresponds to the flow map prediction, please see Fig.~\ref{fig:illustrative_network}(a). However, such an approach fails to guarantee property \textbf{(P1)} by design, and instead, the identity mapping must be learned. Moreover, the input-based derivatives of MLPs are themselves nontrivial neural networks~\cite{sitzmann2020implicit}, and do not permit a distinction between instantaneous velocity \textbf{(P2)} and flow map derivatives of nonzero time span. This presents complications for ensuring a stable composition-based objective (c.f. Eq.~\eqref{eq:composition-detail}).

Rather than use a standard coordinate-based network we propose a 2-tiered network design, please see Fig.~\ref{fig:illustrative_network}(b) for an overview. The first network, which we denote a $f_{\nu} : \mathbb{R}^n \times \mathbb{R} \rightarrow \mathbb{R}^d$ learns a $d$-dimensional spatiotemporal representation of the domain that is tasked with property \textbf{(P2)}, learning a representation of instantaneous velocity. We condition the representation $f_{\nu}$ with a given time span $\tau$, via the following multiplicative scaling:
\begin{equation}
\label{eq:z0}
\mathbf{z}^{(0)} = \sigma_{\nu}(\tau \mathbf{m}^{(0)}) \odot f_{\nu}(\mathbf{x},t),
\end{equation}
where $\mathbf{m}^{(0)} \in \mathbb{R}^d$ is a learnable vector aimed to reconcile the scaling of $\tau$ -- initially expressed in terms of the physical domain -- for the neural representation. The function $\sigma_{\nu}$ is a nonlinearity that serves to squash values into a predetermined range, in practice this is set as a hyperbolic tangent, while $\odot$ indicates element-wise multiplication. The second network, which we denote $f_{\tau} : \mathbb{R}^n \times \mathbb{R} \rightarrow \mathbb{R}^d$, similarly learns a $d$-dimensional spatiotemporal representation but one that is specific to the flow map for nonzero time spans. We combine the two representations, $f_{\nu}$ and $f_{\tau}$, through a residual connection:
\begin{equation}
\label{eq:z1}
\mathbf{z}^{(1)} = \mathbf{z}^{(0)} + \sigma_{\nu}\left(\tau \mathbf{m}^{(1)}\right) \odot \sigma_{\tau}\left(\mathbf{z}^{(0)} \odot (W^{(1)} f_{\tau}(\mathbf{x},t)) \right),
\end{equation}
where $\mathbf{m}^{(1)} \in \mathbb{R}^d$ serves the same purpose as $\mathbf{m}^{(0)}$, and $W^{(1)} \in \mathbb{R}^{d \times d}$ is a learnable linear transformation. 
\new{
A consequence of the above construction is that the derivative w.r.t $\tau$ when $\tau = 0$ evaluates to
\begin{equation}
 \frac{d \mathbf{z}^{(1)}}{d \tau} = \frac{d \mathbf{z}^{(0)}}{d \tau} = \left(\sigma_{\nu}' \mathbf{m}^{(0)}\right) \odot f_{\nu}(\mathbf{x},t).   
\end{equation}
}
Subsequent representations are formed via residual connections, \new{in order to preserve the above derivative}:
\begin{equation}
\label{eq:zl}
\mathbf{z}^{(l)} = \mathbf{z}^{(l-1)} + \sigma_{\nu}\left(\tau \mathbf{m}^{(l)}\right) \odot \sigma_{\tau}\left(W^{(l)} \mathbf{z}^{(l-1)} \right),
\end{equation}
and, finally the last layer $L$ applies a single linear transformation to give us the output position, wherein we also include a skip connection for the input position:
\begin{equation}
\label{eq:zlast}
\hat{\Phi}(\mathbf{x},t,\tau) = \mathbf{x} + W^{(L)} \mathbf{z}^{(L-1)}.
\end{equation}

Returning to our properties, we note that this network design, by construction, satisfies the identity mapping \textbf{(P1)}, so long as the chosen activation function $\sigma_{\nu}$ satisfies $\sigma_{\nu}(0) = 0$. The multiplicative scaling performed at each layer ensures that all representations will be zero vectors throughout the network. Critically, we \emph{do not} introduce bias vectors, in order to guarantee this identity mapping. More importantly, \new{the designed residual connections} lead to a particularly simple network for the flow map derivative at $\tau = 0$ \textbf{(P2)}:
\begin{equation}
\label{eq:zderiv}
\frac{d \hat{\Phi}(\mathbf{x},t,0)}{d \tau} = W^{(L)} (\mathbf{m}^{(0)} \odot f_{\nu}(\mathbf{x},t)).
\end{equation}
Please see the appendix for the supporting derivation. There are two implications of this result. First, instantaneous velocity of the flow map does not depend on the representation $f_{\tau}$, as depicted in Fig.~\ref{fig:illustrative_network}(c); in fact it is entirely decoupled from the rest of the network (c.f. Eqs.~\eqref{eq:z1} and~\eqref{eq:zl}). Hence, we can directly optimize instantaneous velocity of the flow map for the vector field $\boldsymbol{\nu}$ using Eq.~\eqref{eq:zderiv}, without making reference to the remainder of the model. In turn, a flow map that satisfies instantaneous velocity helps ``prime'' the model in satisfying the self-consistency criterion, and ensures stability, e.g. we can choose to freeze the parameters associated with $f_{\nu}$, and $W^{(L)}$ when optimizing Eq.~\eqref{eq:composition-expectation}, and the network's representation of instantaneous velocity will remain unchanged. Secondly, the simplicity of this derivative network ensures that we can easily optimize for the vector field. \new{In contrast, for a standard MLP (c.f. Fig.~\ref{fig:illustrative_network}) its instantaneous velocity would amount to an involved derivative network~\cite{sitzmann2020implicit} to be optimized. This network is no different in structure for $\tau > 0$, and as a consequence, optimizing for both instantaneous velocity, and derivatives for $\tau > 0$, would require a careful balancing act.}

We remark that our network bears similarity to prior work on flow map representations~\cite{bilovs2021neural,han2021exploratory}. In particular, the distinction between spatiotemporal coordinates and time span is considered by Han et al.~\cite{han2021exploratory}, yet the ability to distinguish properties of the flow map for $\tau = 0$ time span is not studied. Our network design is inspired by Bilo{\v{s}} et al.~\cite{bilovs2021neural}, where they similarly consider residual connections. However, we make more precise the role of residual architectures in regards to flow map derivatives, and the relationship with the vector field, this being the only source of supervision in our work.

What remains is a specific instantiation of functions $f_{\nu}$ and $f_{\tau}$. Though in principle these could be arbitrary neural networks, in practice we adapt prior work on learnable feature grids~\cite{takikawa2021neural,weiss2021fast,muller2022instant}, where the parameters of the model are, in part, comprised of learnable spatiotemporal feature grids, each of varying resolution. For each feature grid we perform linear interpolation to obtain a feature vector, and concatenate the vectors obtained across all grids. We then apply a shallow MLP to the concatenated vector. For simplicity, grid cells are accessed exactly, rather than hashed as proposed in M{\"u}ller et al.~\cite{muller2022instant}. The bulk of the network parameters thus lie in the feature grids, rather than MLP weights, and so in practice the dimensionality $d$ need not be too large -- in practice we set $d = 64$. As a result, the cost of evaluating the network is inexpensive, requiring (1) interpolation of feature vectors from a set of grids, and (2) applying several matrix-vector multiplication operations (c.f. Eqs.~\eqref{eq:z0}--\eqref{eq:zlast}).

\subsection{Optimization Scheme}
\label{sec:optimization}

% overview of optimization scheme
Our optimization scheme proceeds in two phases. In the first phase we optimize for the flow map's instantaneous velocity, while in the second phase we optimize for the self-consistency criterion, in order to learn the flow map over the full spatiotemporal domain, and varying time spans.

\textbf{Vector field optimization.} To find the flow map's instantaneous velocity, we minimize the following objective:
\begin{equation}
    \mathcal{L}_{\nu} = \mathbb{E}_{(\mathbf{x},t) \in \mathcal{D}} \left[ \lVert W^{(L)} (\mathbf{m}^{(0)} \odot f_{\nu}(\mathbf{x},t)) - \boldsymbol{\nu}(\mathbf{x},t) \rVert \right].
\end{equation}
This amounts to optimizing over the parameters of $f_{\nu}$, e.g. the multi-level feature grid, shallow MLP, vector $\mathbf{m}^{(0)}$, and final projection $W^{(L)}$. We emphasize that it is only this phase of optimization that requires the vector field for supervision. The relevant portion of the model (c.f. Fig.~\ref{fig:illustrative_network}(c)) can encode the vector field in a persistent manner, even as we optimize for the flow map in the subsequent phase, and thus we may discard the vector field post optimization. For large-scale vector fields that may not fit in memory, this gives us the opportunity to learn a compressed vector field representation, e.g. one that can fit in memory for use in the next optimization phase, as well as at inference time.

\begin{figure}
  \centering
 	\includegraphics[width=1\linewidth]{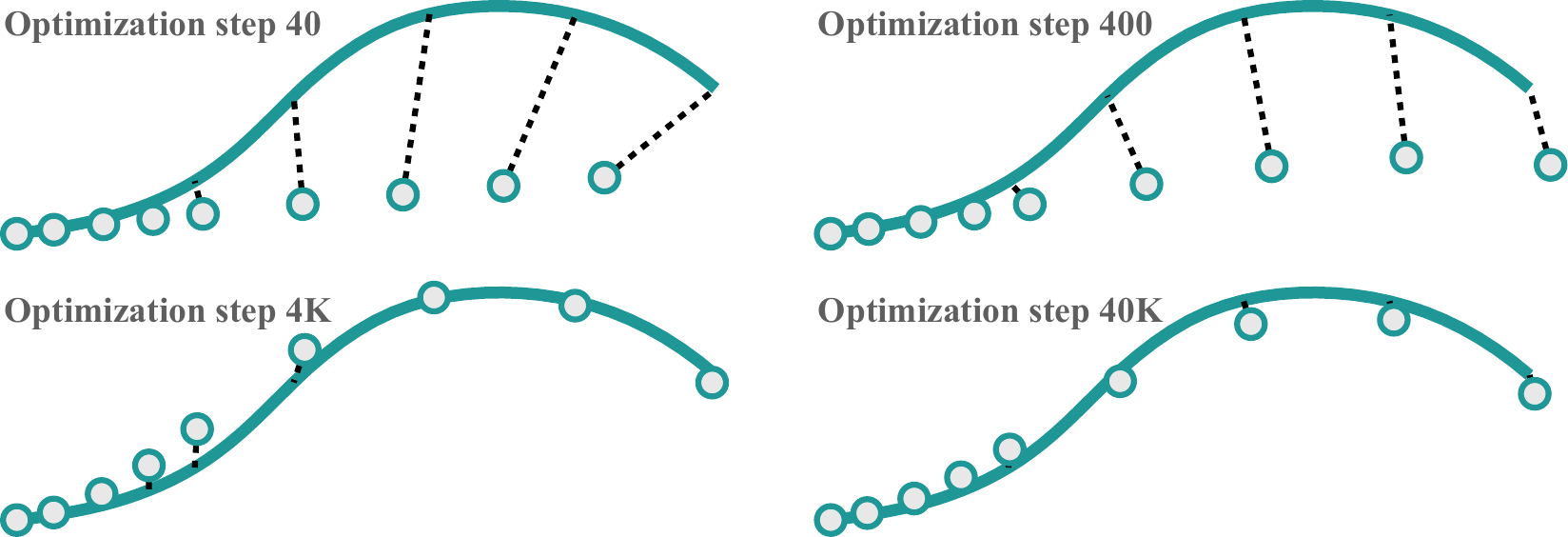}
	\caption{We illustrate how flow map optimization incrementally learns longer time spans over the course of optimization. Initially (40 steps), the flow map provides us with a linear approximation, owing to its instantaneous velocity well-representing the vector field. Over optimization, the flow map becomes more accurate in its predictions for longer time spans. }
    \label{fig:optim}
\end{figure}

\textbf{Flow map optimization.} To learn the flow map, we are guided by the proposed self-consistency criterion of Eq.~\eqref{eq:composition-detail}. Although we often find taking just a single step is sufficient for giving accurate flow maps, for certain datasets, we find it useful to instead take multiple steps in optimizing this loss. More specifically, we define $\bar{\Phi}(\mathbf{x},t,\varepsilon) = (\hat{\Phi}(\mathbf{x},t,\varepsilon),t+\varepsilon,\varepsilon)$ to be the resulting position, subsequent time step, and time span $\varepsilon$, from applying the flow map. Then for some target time span $\tau$, we can compose the flow map into multiple steps $k \in \mathbb{Z}^{+}$ as follows:
\begin{equation}
\label{eq:steps}
\hat{\Phi}_k(\mathbf{x},t,\tau) = \underbrace{\bar{\Phi} \, \circ \, \bar{\Phi} \, \circ \, \cdots \, \circ}_{k-1} \bar{\Phi}(\mathbf{x},t,\frac{\tau}{k}).
\end{equation}
We then redefine our self-consistency loss as:
\begin{equation}
    \label{eq:comp-loss}
    l_s(\mathbf{x},t,\tau) = \bigg\lVert \frac{\partial \hat{\Phi}(\mathbf{x}, t, \tau)}{\partial \tau} - \frac{\partial \hat{\Phi}(\hat{\Phi}_k(\mathbf{x}, t, \tau),t+\tau,0)}{\partial \tau} \bigg\rVert.
\end{equation}

Experimentally, we find that the number of steps $k$ to take can be set in proportion to the time span. In particular, for a given time span $\tau$ if we let $\tau_g$ be this value's expression in grid units of the field's time domain, then we find it sufficient to set $k(\tau) = \lceil \sqrt{\tau_g} \rceil$. Given that our network architecture permits efficient evaluation, this is reasonably cheap to compute, especially in relation to full numerical integration. For further efficiency, we find that the right-hand side of Eq.~\eqref{eq:comp-loss} can be frozen during optimization, and hence we only optimize for the flow map derivative under nonzero time span $\tau$.

During flow map optimization, we may freeze the network dedicated to instantaneous velocity, if not fine-tune it with a learning rate much smaller than the flow map portion of the network, e.g. approximately 2 orders of magnitude less. \new{Moreover, $\tau$-scaled residual connections ensure that the derivative network for nonzero $\tau$ is not overly complex, e.g. for small $\tau$ it will remain close to the derivative at $\tau = 0$.} We find the ability of the flow map to well-represent small time spans (c.f. Eq.~\eqref{eq:zero_advect}) significantly helps stabilize optimization, and in practice, we find that the flow map incrementally improves on larger time spans over the course of optimization, please see Fig.~\ref{fig:optim} for an illustration.
\section{Results}
In this section we experimentally evaluate our method -- herein termed NIFM for Neural Integration-free Flow Maps -- for both 2D and 3D time-varying vector fields, comparing against various baselines that accelerate flow map computation in different ways. \new{A requirement that is common to all baselines is access to samples of the flow map. Unless otherwise stated (c.f. Sec.~\ref{subsec:error}), the methods against which we compare NIFM are based on flow maps generated via $4^{th}$ order Runge-Kutta integration (RK4), with step size set to half of the temporal voxel size. We also use this very integration scheme to generate ground-truth flow map samples for the purposes of evaluation.} In Table~\ref{tab:datasets} we list the datasets used for comparison purposes. Further, all reported computational timings are based on a system with 12-core CPU AMD Ryzen 9 3900X, 16GB RAM, and GPU NVIDIA GeForce RTX 2080 Ti with 12GB memory.

We consider the flow map super resolution technique proposed by Jakob et al.~\cite{jakob2020fluid}, wherein we train a convolutional neural network (CNN) model using the 2D fluid flow dataset provided by the authors. To train the CNN we generate 16x downsampled flow maps along with their corresponding high-resolution ground truth flow maps, varying start times and time span of the integration, to permit model generalization for arbitrary start time/duration.

Additionally, we compare our method with the deep learning based Lagrangian interpolation technique proposed by Han et al.~\cite{han2021exploratory}. This technique uses an encoder-decoder network and is most similar to ours in terms of the input data the model expects, and the output of the model. We train the model on flow map samples computed by, first, generating seeds sampled uniformly at random in space and time, and secondly, integrating for varying small time spans. This flow map sampling technique is intended to resemble the Lagrangian short generation scheme proposed by the authors. We made a minor modification to the network by removing the ReLU activation function used in the output layer, allowing the model to output negative values. Further, we compare our method with a SIREN~\cite{sitzmann2020implicit} that tacks time span on as an additional coordinate, along with particle space-time coordinates (c.f. Fig.~\ref{fig:illustrative_network}(a)). We train the SIREN with the same data used to train the encoder-decoder model. Note that we could use a hybrid grid-MLP model~\cite{muller2022instant,weiss2021fast} in lieu of a standard coordinate-based MLP, but for 3D unsteady flows this would require storage of a 5D grid, which is not feasible.

We also compare our method against the recent work by Li et al.~\cite{li2022efficient}, where the authors showed an improvement over prior work in efficiently interpolating Lagrangian representation to obtain new trajectories. \new{Note that the representation of flow in our datasets is Eulerian, whereas Li et al. works with particle-based data, thus, requiring a conversion from the former to the latter. For a fair comparison, we convert the Eulerian representation into a Lagrangian one by first placing $n_s$ number of seeds in the domain uniformly at random, where $n_s$ is the spatial resolution of the vector field data, and integrate these seed points via RK4. The temporal frequency with which we store particle positions is set as the temporal resolution of the field. Furthermore, the Lagrangian representation is limited to the temporal duration on which we are evaluating, to have a better distribution of particles throughout the domain}.

Last, we compare our method with the streakline vector field (SVF) work of Weinkauf et al.~\cite{weinkauf2010streak}. \new{Specifically, the SVF is first precomputed by estimating flow map derivatives, computed via RK4, and then at runtime streaklines are generated by integrating the SVF. We view this as a fair comparison to our technique in that both approaches incur a precomputation cost, and thus we aim to compare the computation and storage requirement for the representations, as well as the accuracy and computation efficiency for generating streaklines.}

\begin{table}[]
\caption{We list all datasets and their respective sizes used in experiments.}
\label{tab:datasets}
\centering
\scalebox{0.8}{
% \begin{tabular}{|c|c|}
% \hline
% Dataset       & Res (t,x,y(,z)) \\ \hline
% Double Gyre   & 500x400x200     \\ \hline
% Cylinder      & 1001x400x50     \\ \hline
% Boussinesq    & 2001x450x150    \\ \hline
% Tornado       & 50x128x128x128  \\ \hline
% Scalar Flow   & 151x100x178x100 \\ \hline
% Half-Cylinder & 151x640x240x80  \\ \hline                 
% \end{tabular}}
% \end{table}
\begin{tabular}{cc}
\hline
Dataset       & Res {[}t,x,y(,z){]} \\ \hline
Double Gyre   & 500x400x200         \\
Cylinder      & 1001x400x50         \\
Boussinesq    & 2001x450x150        \\
Fluid Simulation & 1001x512x512     \\
Tornado       & 50x128x128x128      \\
Scalar Flow   & 151x100x178x100     \\
Half-Cylinder & 151x640x240x80   
\end{tabular}}
\end{table}

\begin{figure*}[t]
\centering
\includegraphics[width=1\linewidth]{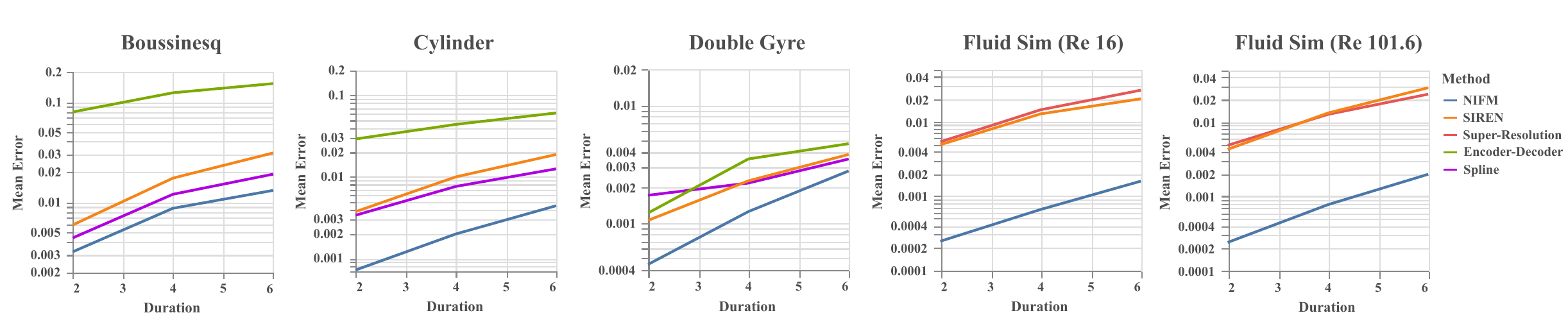}
\caption{We show the quantitative evaluation of flow map approximation methods across different datasets, and across different time spans, beginning at start times for which flow features have largely resolved.} 
\label{fig:quantitative}
\end{figure*}

\subsection{Implementation details}
We first describe the details of our network architecture, followed by details on optimization.

\textbf{Network architecture settings}
The design of $f_{\nu}$ and $f_{\tau}$ rely on parameter settings related to the multi-level feature grid, as well as the MLP. The feature grids for $f_{\nu}$ and $f_{\tau}$ are of identical design, where we use a 4-level feature grid, and each level is of a different spatial resolution. Specifically, for a given axis of resolution $w$ at level $l$, we set the resolution at the next level to be $w^{s \cdot l}$, with resolution scaling factor $s$ set to 1.65, following the guidance of M{\"u}ller et al.~\cite{muller2022instant}. Each grid stores $8$-dimensional feature vectors at its nodes, and thus the resulting concatenated feature is $32$-dimensional. We employ 2 and 1-layer MLPs for $f_{\nu}$ and $f_{\tau}$, respectively, along with activation $\sigma_{\tau}$ chosen to be a Swish activation~\cite{hayou2018selection}. Experimentally we found Swish to outperform other more standard activations for INRs, e.g. ReLU, sin, consistent with findings in AutoInt~\cite{lindell2021autoint}. We control for the size of the network by a compression ratio, expressed as the ratio of the vector field size to the network size. We adjust the spatial resolution of the feature grids to best match a provided compression ratio, but leave the MLPs unchanged as they comprise a tiny portion of the model. Last, we use a 3-layer MLP with $64$ layer width for the residual network. Unless otherwise specified, we use a compression ratio of $10$ for all 2D datsets, and customize compression ratios for 3D as appropriate.
 
\begin{table}[]
\caption{We report the preprocessing times for different methods across 2D unsteady flows, along with corresponding timings for FTLE computation, varying time span and image resolution.}
\label{tab:time}
\centering
\scalebox{0.7}{%
    \begin{tabular}{|c|c|c|c|c|c|c|c|}
    \hline
    Dataset                      & FTLE res           & $\tau$ & \begin{tabular}[c]{@{}c@{}}Inference \\time(s)\end{tabular} & \begin{tabular}[c]{@{}c@{}}Preprocessing \\time(min)\end{tabular} & CR & \begin{tabular}[c]{@{}c@{}}Storage \\(MB)\end{tabular} &method       \\ \hline
    \multirow{4}{*}{Fluid Sim} & \multirow{4}{*}{512x512}  & \multirow{4}{*}{7}   & 21.161                                                 & -              & -       &   2003 & GT \\ \cline{4-8} 
                                      &                           &                      & \textbf{0.585}                                & 48.01             & 10     &   189   & NIFM         \\ \cline{4-8} 
                                      &                           &                      & 2.010                                        & 63.33              & 1       &              & Siren   \\ \cline{4-8} 
                                      &                           &                      & 4.701                                          & 1104.60           & -       &              & FSR \\ \hline
    \multirow{4}{*}{Cylinder}         & \multirow{4}{*}{1200x150}  & \multirow{4}{*}{1}   & 1.853                                         & -                & -         &     153     & GT \\ \cline{4-8} 
                                      &                           &                      &  \textbf{0.055}                              & 33.50              & 10         &     16     & NIFM         \\ \cline{4-8}
                                      &                           &                      & 0.554                                        & 74.26              & -           &          & ED   \\ \cline{4-8} 
                                      &                           &                      & 0.324                                        & 41.76              & 1            &        & Siren   \\ \cline{4-8} 
                                      &                           &                      & 29.94                                              & 0.04         & -             &       & Spline    \\ \hline
    \multirow{4}{*}{Boussinesq}       & \multirow{4}{*}{450x1350} & \multirow{4}{*}{0.5} & 2.220                                          & -                 & -            &  1030  & GT \\ \cline{4-8} 
                                      &                           &                      &  \textbf{0.079}                               & 37.25             & 10            &      97    & NIFM         \\ \cline{4-8} 
                                      &                           &                      & 0.938                                        & 122.89              & -             &        & ED   \\ \cline{4-8} 
                                      &                           &                      & 0.621                                       & 63.28                & 1       &             & Siren   \\ \cline{4-8} 
                                      &                           &                      & 91.57                                             & 0.15            & -       &       & Spline    \\ \hline
    \multirow{4}{*}{\begin{tabular}[c]{@{}l@{}}Double Gyre\end{tabular}}      & \multirow{4}{*}{1200x600}  & \multirow{4}{*}{10}  & 34.453                      & -        & -   &  611   & GT \\ \cline{4-8} 
                                      &                           &                      &  \textbf{1.020}                               & 34.80                 & 10        &      29     & NIFM         \\ \cline{4-8}
                                      &                           &                      & 6.278                                        & 40.70                  & -         &         & ED   \\ \cline{4-8}
                                      &                           &                      & 1.689                                        & 19.84                 & 1       &    & Siren   \\ \cline{4-8} 
                                      &                           &                      & 252.63                                             & 0.29            & -        &          & Spline    \\ \hline
    \end{tabular}%
}
\end{table}
 
\textbf{Optimization details}
For both phases of optimization we use Adam~\cite{kingma2015adam}, where we take a total of $40,000$ optimization steps and decay the learning rate every $8,000$ steps. Specific to optimization phase, in fitting to the vector field we use a learning rate of $0.02$, while for flow map optimization we use a learning rate of $0.01$ -- fitting the flow map derivative to the vector field is quite stable, and benefits from larger learning rates. In optimizing for the flow map, we have the choice of leaving the instantaneous velocity portion of the network frozen, or fine-tuning its weights to compensate for the remainder of the network. Although we find that both give results of comparable accuracy, in some occasions we found that fine-tuning can mitigate small grid-based artifacts in the output when leaving these weights frozen, and hence we fine-tune this portion of the network, using a learning rate of $0.0008$.

Recall that our method supports a maximum time span $\tau_{max}$ on which to sample during optimization. Though in principle we could optimize for the full time span of a given dataset, we find that performance can suffer, especially for datasets exhibiting complex temporal dynamics. Thus, as a compromise we set a limit on $\tau_{max}$ during optimization, and at inference time, for any target $\tau > \tau_{max}$ we take multiple steps with our network until reaching the desired span $\tau$. Specifically, for all 2D datasets, expressed in terms of grid units we set $\tau_{max} = 48$ unless otherwise specified. For 3D datasets we customize $\tau_{max}$ based on grid resolution, and complexity of the flows.

\begin{figure*}[t]
    \centering
    \includegraphics[width=1\linewidth]{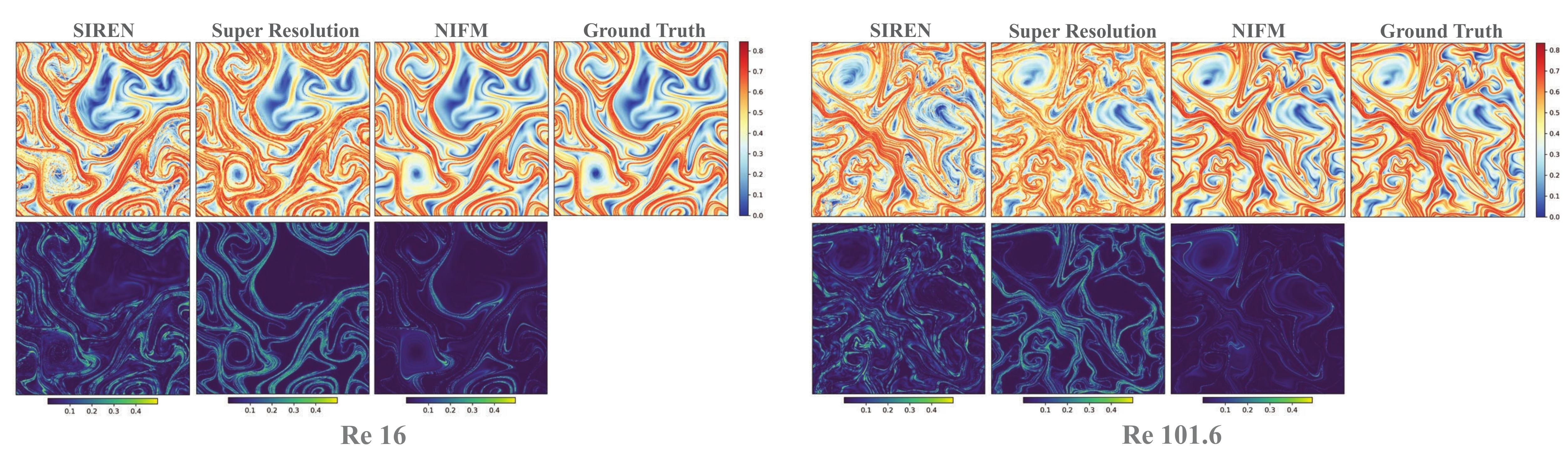}
    \caption{We compare FTLE (top row) and integration error (bottom row) for two Fluid Simulation datasets (Re 16 and Re 101.6) across different baselines. The left column corresponds to particles integrated beginning at $t_0 = 0$ for duration $\tau=7$, while the right column corresponds to particles integrated starting at $t_0 = 2$ and $\tau = 7$.} 
    \label{fig:ftle_fluid}
\end{figure*}

\begin{figure*}[t]
    \centering
    \includegraphics[width=1\linewidth]{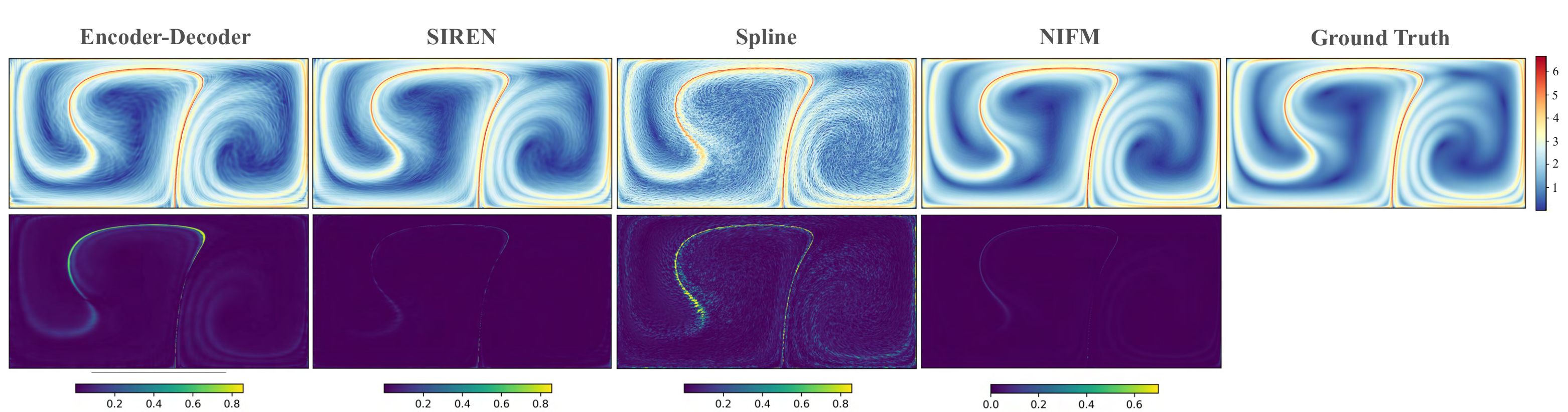}
    \caption{We compare FTLE (top row) and integration error (bottom row) for different baselines for the Double Gyre dataset. Particles are integrated from $t_0=0$ for a time-span $\tau=10$.} 
    \label{fig:ftle_dg}
\end{figure*}

\subsection{2D unsteady flow}

We first conduct experimental comparisons for various 2D time-varying flow fields. Specifically, we evaluate different techniques by computing the error in flow map approximations over varying seed points (spatial position and starting time) that have been integrated for varying time spans. We express error as the averaged Euclidean distance between the ground-truth flow map output, and the approximation scheme's output, normalized by the domain's bounding-box diagonal length. In Fig.~\ref{fig:quantitative} we present quantitative results comparing our method against different baselines, and in Table~\ref{tab:time} we report inference and preprocessing times. Specifically, for the pathline interpolation approach of Li et al.~\cite{li2022efficient}, preprocessing refers to the time required to fit B-splines, while for Jakob et al.~\cite{jakob2020fluid} this refers to the time required to optimize the CNN for super resolution. For all remaining methods, preprocessing refers to the time required for optimizing to an individual flow field.

\begin{figure}[!t]
    \centering
    \includegraphics[width=1.0\linewidth]{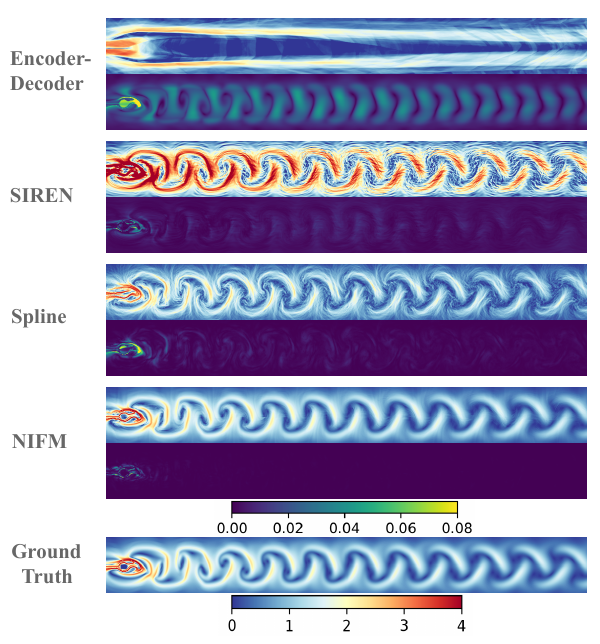}
    \caption{We show the FTLE (top of each pair) and error maps (bottom of each pair) for the flow over cylinder dataset generated by integrating particles starting at $t_0=18$ for a time-span $\tau=1$.} 
    \label{fig:ftle_cy}
\end{figure}

\begin{figure}[!t]
    \centering
    \includegraphics[width=1.0\linewidth]{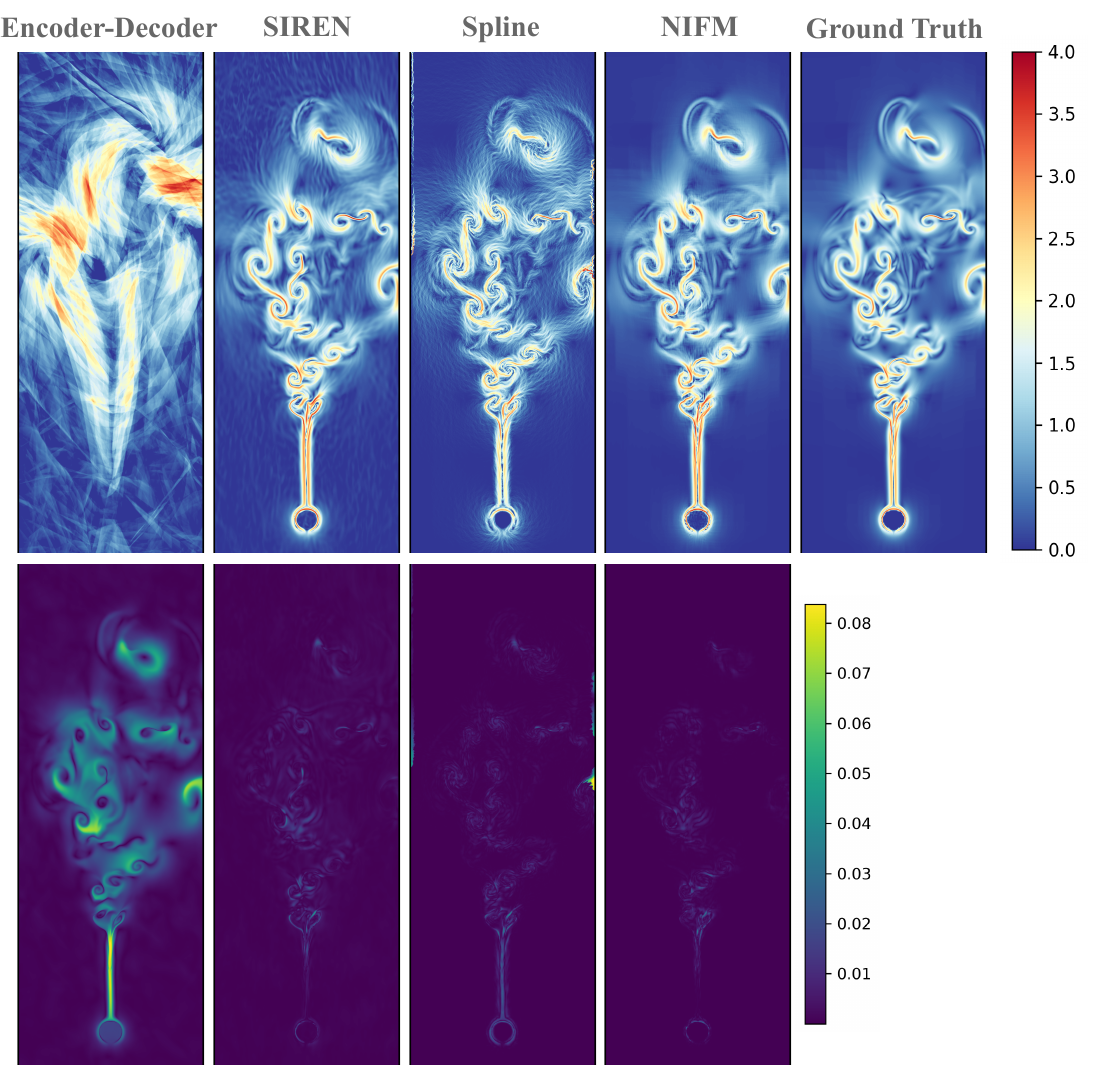}
    \caption{We show the FTLE (top of each pair) and error maps (bottom of each pair) for the Boussinesq dataset generated by integrating particles starting at $t_0=11.3$ for a time-span $\tau=0.5$.} 
    \label{fig:ftle_bo}
\end{figure}

In comparing the fluid simulation flows of varying Reynolds numbers, we find that our method sees consistent improvement in accuracy over SIREN and super resolution, while achieving faster inference times. We note that the super resolution approach requires optimizing a CNN over a collection of flow maps just once, and thus can generalize to low-resolution flow maps at inference time, albeit restricted to flows resembling those observed during training. Our method is limited to just a single dataset at a time, but nevertheless, our training times scale well in terms of standard INRs (e.g. SIREN), while exhibiting faster inference and more accurate flow map approximations. Qualititative results for the fluid simulation flows are shown in Fig.~\ref{fig:ftle_fluid} in the form of the FTLE -- \new{computed using the method of Haller~\cite{haller2001lagrangian}} -- and color-encoded flow map errors. For high Reynolds number flows, we see that the super resolution method can fail to adapt to the rate at which particles separate, as indicated by the color shift, while also blurring out detailed ridges in the FTLE.  Our method, however, excels in capturing FTLE ridges, while remaining efficient to compute, since the super resolution method still requires computing a low-resolution flow map as input to a (otherwise highly efficient) CNN. Recall that our method employs a compression ratio of $10$ for all 2D experiments, which limits the grid resolution, and thus might limit the details we can reproduce in the flow map. However, from these results, we see that the coarser feature grid resolution does not limit the spatial resolution of the FTLE.

\begin{figure*}[t]
\centering
\includegraphics[width=1\linewidth]{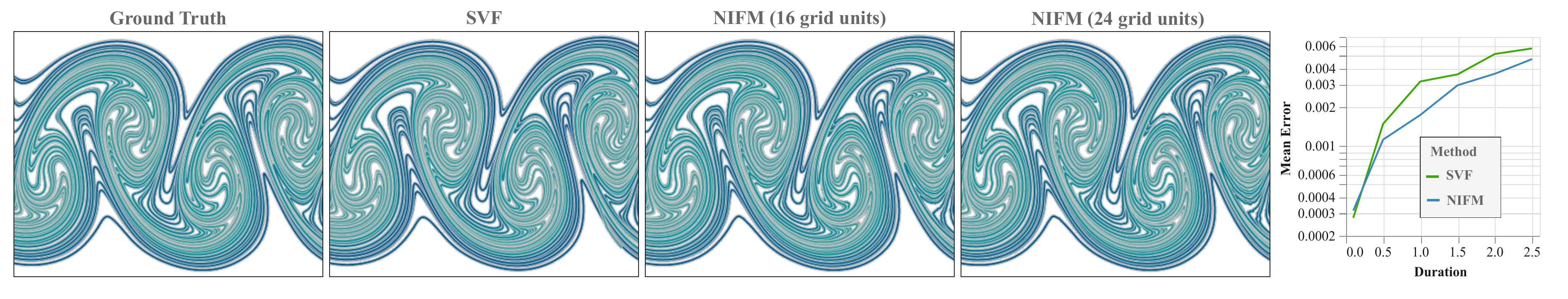}
\caption{We compare our method's ability to compute streaklines against the streakline vector field technique~\cite{weinkauf2010streak}, which only necessitates integrating a derived vector field. Qualitatively and quantitatively we find that our method produces comparable results, where we show varying step sizes used for evaluating the flow map.}
\label{fig:streaklines}
\end{figure*}

In comparing our method to other baselines (c.f. Fig.~\ref{fig:quantitative}) for Double Gyre, Cylinder, and Boussinesq, we find that our method obtains higher accuracy in relation to other techniques. Prior INR methods such as the encoder-decoder architecture of Han et al.~\cite{han2021exploratory}, or a pure coordinate-based approach~\cite{sitzmann2020implicit} poorly generalize. We find that for small step sizes, the performance of these methods in fact steeply declines, as numerical error accumulates with the more steps taken. We attribute this to the basic limitations of the network architectures employed, failing to address the properties (identity mapping, instantaneous velocity) we target in our network design. The inability to generalize in these methods is further demonstrated qualitatively for Figs.~\ref{fig:ftle_dg} - \ref{fig:ftle_bo}. \new{Pathline interpolation~\cite{li2022efficient} is notable in its small precomputation cost. Nevertheless, the method is less accurate in preserving the flow map, while incurring a high computation cost at runtime.}

We additionally evaluate our technique both quantitatively and qualitatively for the computation of streaklines. In Fig.~\ref{fig:streaklines} we show streaklines for the Cylinder dataset. We compare our method with SVF~\cite{weinkauf2010streak}. We can see that both the techniques are able to capture the vortices of the dataset faithfully, and are visually indistinguishable from the ground truth streaklines. Quantitatively both the techniques consistently incur low streakline error staying within the margin of $10^{-3}$ magnitude (relative to the bounding box diagonal). Interestingly, we find that both methods have comparable inference time as well, as reported in Table~\ref{tab:streaklines}, despite the fact the streakline vector field evaluates its field fewer times than our neural flow map, since we must take multiple steps for sufficiently long time spans. However, an advantage of our method lies in data parallelism; we can evaluate the flow map over varying space/time/duration in a single batch, whereas integrating the streakline vector field is, by necessity, a sequential process. We further note that SVF precomputation is quite expensive, both in terms of speed and storage space. In Table~\ref{tab:streaklines} we can see that the computation of the entire 4D SVF has very large storage requirements (160GB), whereas our method is in proportion to the size of the vector field (77MB). We note that while our technique can be easily scaled to 3D datasets, SVF preprocessing for 3D unsteady flows is infeasible in practice, necessitating a 5D grid for storage.

\begin{table}[!t]
\caption{We report storage requirements, preprocessing time and inference time for computing streaklines on the Cylinder dataset, comparing our method against the streakline vector field technique~\cite{weinkauf2010streak}.}
\label{tab:streaklines}
\centering
\scalebox{0.9}{
\begin{tabular}{|c|c|c|c|}
\hline
Method  & \begin{tabular}[c]{@{}c@{}}Preprocessing\\ Time\\ (min)\end{tabular} & \begin{tabular}[c]{@{}c@{}}Inference\\ Time\\ (sec)\end{tabular} & \begin{tabular}[c]{@{}c@{}}Storage\\ \end{tabular} \\ \hline
Ground Truth & NA  & 21.391 & 160.20 MB \\  \hline
SVF & 130.407 & 1.204 & 160.36 GB \\ \hline
NIFM (16 grid steps) & \multirow{2}{*}{40.060} & 0.952 & \multirow{2}{*}{77.20 MB}\\ \cline{1-1} \cline{3-3}
NIFM (24 grid steps) & &0.671 & \\ \hline
\end{tabular}}
\end{table}

\begin{table}[!t]
\caption{We report the processing times as well the FTLE computation times for different method across different 3D unsteady flow datasets.}
\label{tab:3d_ftle_times}
\centering
\scalebox{0.8}{
\begin{tabular}{|c|c|c|c|c|c|c|}
\hline
Dataset                                        & FTLE res                     & $\tau$                                     & \multicolumn{1}{l|}{\begin{tabular}[c]{@{}l@{}}Inference\\ times (s)\end{tabular}} & \multicolumn{1}{l|}{\begin{tabular}[c]{@{}l@{}}Processing\\ times(m)\end{tabular}} & CR & \multicolumn{1}{l|}{Method} \\ \hline
\multicolumn{1}{|c|}{\multirow{4}{*}{Tornado}} & \multirow{4}{*}{128x128x128} & \multirow{4}{*}{50}                     & 27.16                                                                              & -                                                                                 & -  & GT                          \\ \cline{4-7} 
\multicolumn{1}{|c|}{}                         &                              &                                         & \textbf{3.60}                                                                               & 35.55                                                                       & 10        & NIFM                        \\ \cline{4-7} 
\multicolumn{1}{|c|}{}                         &                              &                                         & 14.14                                                                              & 93.21                                                                           & 10   & SIREN                       \\ \cline{4-7} 
\multicolumn{1}{|c|}{}                         &                              &                                         & 286.29                                                                             & 0.87
& - & Spline                      \\ \hline
\multirow{4}{*}{Scalar Flow}                   & \multirow{4}{*}{100x178x100} & \multirow{4}{*}{2.5}                    & 81.72                                                                              & -                                                                              & -     & GT                          \\ \cline{4-7} 
                                               &                              &                                         & \textbf{2.55}                                                                               & 41.66                                                                          & 10     & NIFM                        \\ \cline{4-7} 
                                               &                              &                                         & 21.48                                                                              & 95.57                                                                      & 10         & SIREN                       \\ \cline{4-7} 
                                               &                              &                                         & 291.39                                                                             & 0.81                                                                        & -        & Spline                      \\ \hline
\multirow{3}{*}{Half-Cylinder}                 & \multirow{3}{*}{640x240x80}  & \multicolumn{1}{c|}{\multirow{3}{*}{2}} & 137.41                                                                             & -                                                                                & -   & GT                          \\ \cline{4-7} 
                                               &                              & \multicolumn{1}{c|}{}                   & \textbf{3.82}                                                                               & 45.56                                                                       & 40        & NIFM                        \\ \cline{4-7} 
                                               &                              & \multicolumn{1}{c|}{}                   & 53.52                                                                              & 103.13                                                                        & 40      & SIREN                       \\ \hline
\end{tabular}}
\end{table}

\begin{figure*}[t]
    \centering
    \includegraphics[width=0.9\linewidth]{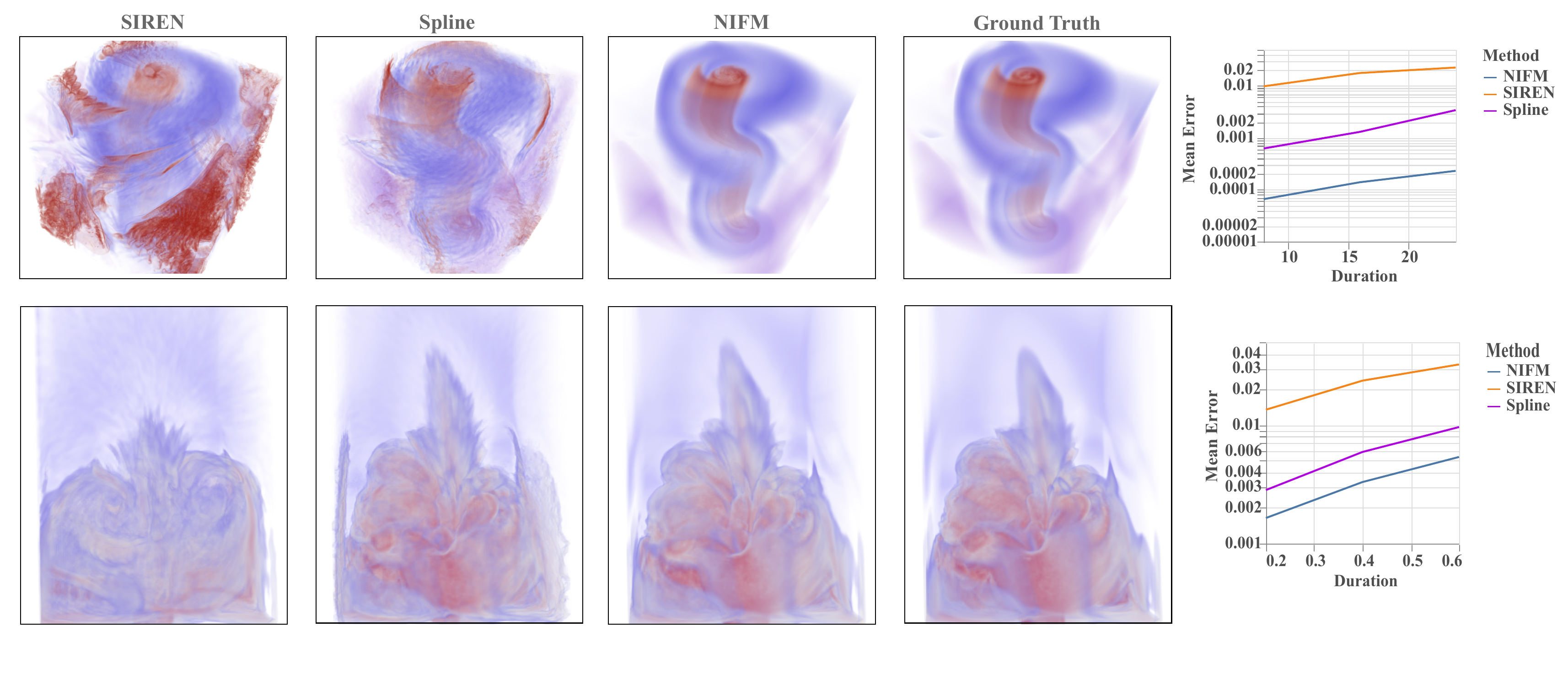}
    \caption{We compare, both qualitatively (volume rendering of FTLE field) and quantitatively (flow map evaluation), our method with standard coordinate-based networks~\cite{sitzmann2020implicit} as well as pathline interpolation techniques~\cite{li2022efficient} for modeling the flow map in 3D unsteady flows. We find our method is quantitatively an improvement over other methods, and qualitatively our method contains fewer visual artifacts.}
    \label{fig:ftle_scalar}
\end{figure*}

\begin{figure*}[t]
    \centering
    \includegraphics[width=0.9\linewidth]{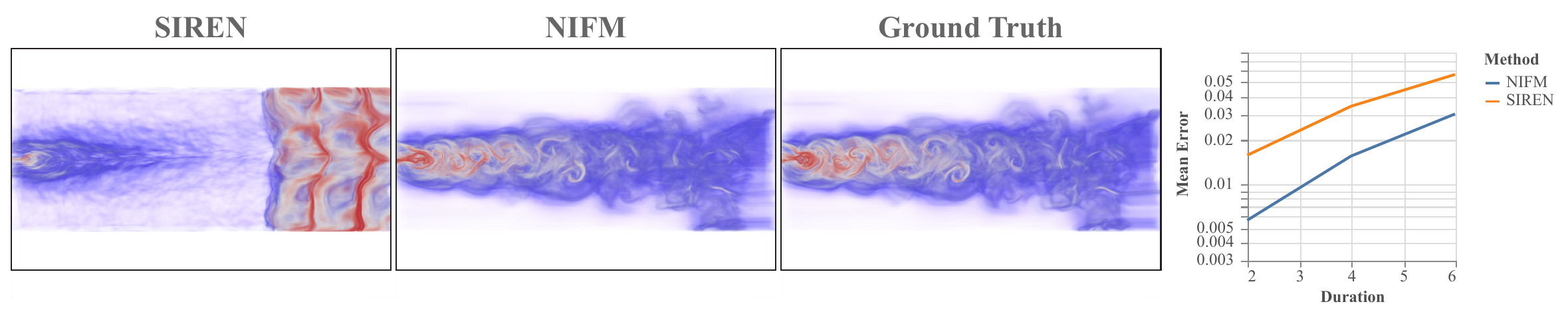}
    \caption{In this figure, we compare our method both quantitatively and qualitatively against SIREN for the Half-Cylinder dataset. We find that our method is able to scale reasonably well to this large dataset, whereas, the SIREN fails to learn meaningful flow maps as can be seen from the FTLE.}
    \label{fig:ftle_half_cylinder}
\end{figure*}

\subsection{3D unsteady flow}

We next evaluate our method on a set of 3D unsteady flows, comparing our method with a SIREN-based flow map~\cite{sitzmann2020implicit} as well as the B-spline pathline interpolation technique~\cite{li2022efficient}. We first compare to the Tornado and Scalar Flow datasets, where we set the $\tau_{max}$ to $8$ and $24$, respectively, to match the temporal complexity in the flows. Fig.~\ref{fig:ftle_scalar} shows qualitative results, via volume-rendering of the FTLE, as well as quantitative results. Our method is an improvement, if not comparable, to prior methods, but we obtain significant gains in inference time, as reported in Table~\ref{tab:3d_ftle_times}. We further compare to the Half Cylinder dataset, a large-scale unsteady flow dataset that cannot be readily stored in memory. We found the pathline interpolation method~\cite{li2022efficient} failed to fit to the data, and thus we limit our comparison to SIREN, please see Fig.~\ref{fig:ftle_half_cylinder}. In this experiment we set $\tau_{max} = 8$ and the compression ratio to $40$ to compensate for the larger data size. We find our method captures turbulent features in the wake of the half-cylinder object ($Re=320$), whereas SIREN faces difficulties in accurately modeling the data. Notably, for this dataset we find our training scheme scales well (c.f. Table~\ref{tab:3d_ftle_times}) relative to the 2D unsteady flow datasets, whereas SIREN's increase in model size leads to slower training times.

\vspace{-.9em}

\begin{figure}[!t]
    \centering
    \includegraphics[width=1.0\linewidth]{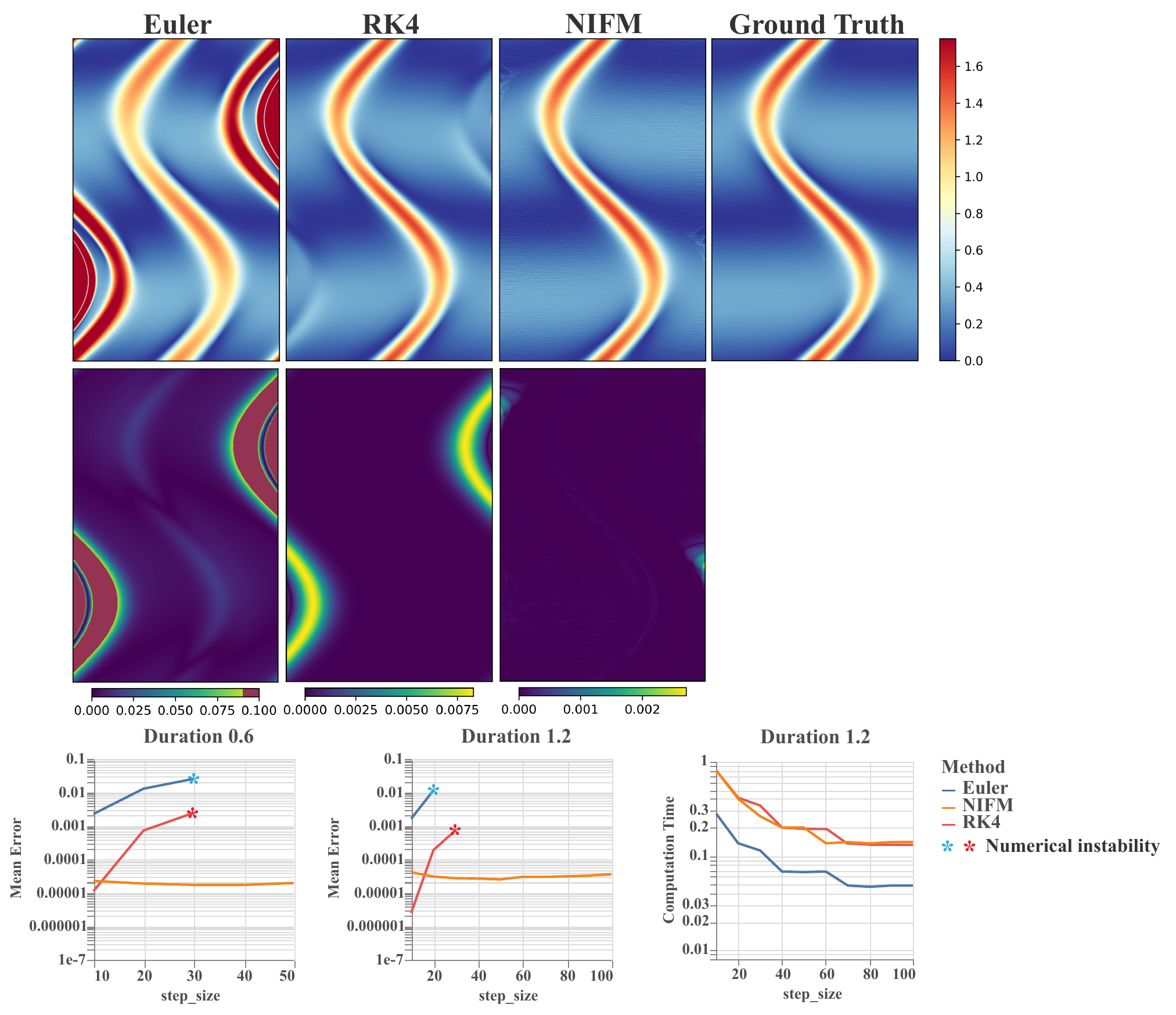}
    \caption{We compare NIFM to Euler and RK4 integration schemes, showing FTLE (top), flow map error (middle), and quantitative evaluation (bottom). We find NIFM performs consistently well across step sizes as compared to Euler and RK4 which can become numerically unstable.} 
    \label{fig:analytical}
\end{figure}

\new{
\subsection{Error analysis: numerical integration}
\label{subsec:error}
Our method can be viewed as a novel technique for integrating a vector field, and thus, it is worth asking: how does our method compare to conventional numerical integration schemes? To help answer this question, we compare NIFM to existing numerical schemes, namely Euler and RK4, evaluated under varying step sizes. For the purpose of evaluation we use the Sine Ridge dataset provided by Kuhn et al.~\cite{kuhn2012benchmark} - as this is a steady flow we adapt our method accordingly. The dataset has an analytically-defined flow map that allows us to compute the flow map error across different schemes. In Fig.~\ref{fig:analytical}, we show the FTLE (first row) and the flow map error (second row) for Euler, RK4, and NIFM. The FTLE is computed for a duration $\tau=1.2$ with step size set to 30, where a single step amounts to 0.01 in the physical domain. We can see that NIFM best captures the FTLE, while maintaining low error in the flow map, in contrast with Euler and RK4. This provides evidence that our method is not merely a fixed linear (e.g. Euler), or higher-order (e.g. RK4) integration scheme, but rather adapts to the features of the data. We further show quantitative results for duration 0.6 and 1.2, again varying the step size. We can see that while NIFM has a consistent performance across all step sizes, the flow map error increases significantly for both RK4 and Euler with increasing step size.
}

\subsection{Ablation: compression and supervision}

Last, we run model ablations to study the effects of various design choices. Due to space limitations we limit ablation to compression, as well as the role of supervision in learning flow maps. Further experiments regarding the architecture choices (number of levels in the multiresolution grid) and optimization scheme (number of steps to take, c.f. Eq.~\ref{eq:steps}) are detailed in the appendix.

In Fig.~\ref{fig:grid_artifact} we show the results of our model, for the FTLE of the Boussinesq, optimized under varying compression ratios. In this experiment we specifically wish to study how compression might impart visual artifacts in derived quantities of the flow map approximation, as a higher level of compression results in coarser feature grids. Indeed, we find that lower levels of compression lead to fewer grid-like artifacts in the resulting FTLE when taking a smaller steps, e.g. in this setting, a step size of 48 grid units in time amounts to an evaluation of the model just 3 times per position. We further report inference times for the smallest and largest level of compressions, and as expected, a larger number of steps requires longer inference times (e.g. more feedforward passes with the network). Interestingly, we find the inference time is fairly consistent across these compression ratios, indicating that the increased resolution of the grid has a negligible impact on this matter. As detailed in the appendix, we also find that the flow map accuracy takes just a small hit in performance across compression ratios, indicating that flow map accuracy might not be predictive of visual artifacts in derived quantities. Nevertheless, as shown in the figure, training times come at a cost with smaller compression ratios. We thus see natural trade-offs in the (1) flow map quality, (2) inference time (hinging on step size), and (3) training time.

\begin{figure}[!t]
    \centering
    \includegraphics[width=0.9\linewidth]{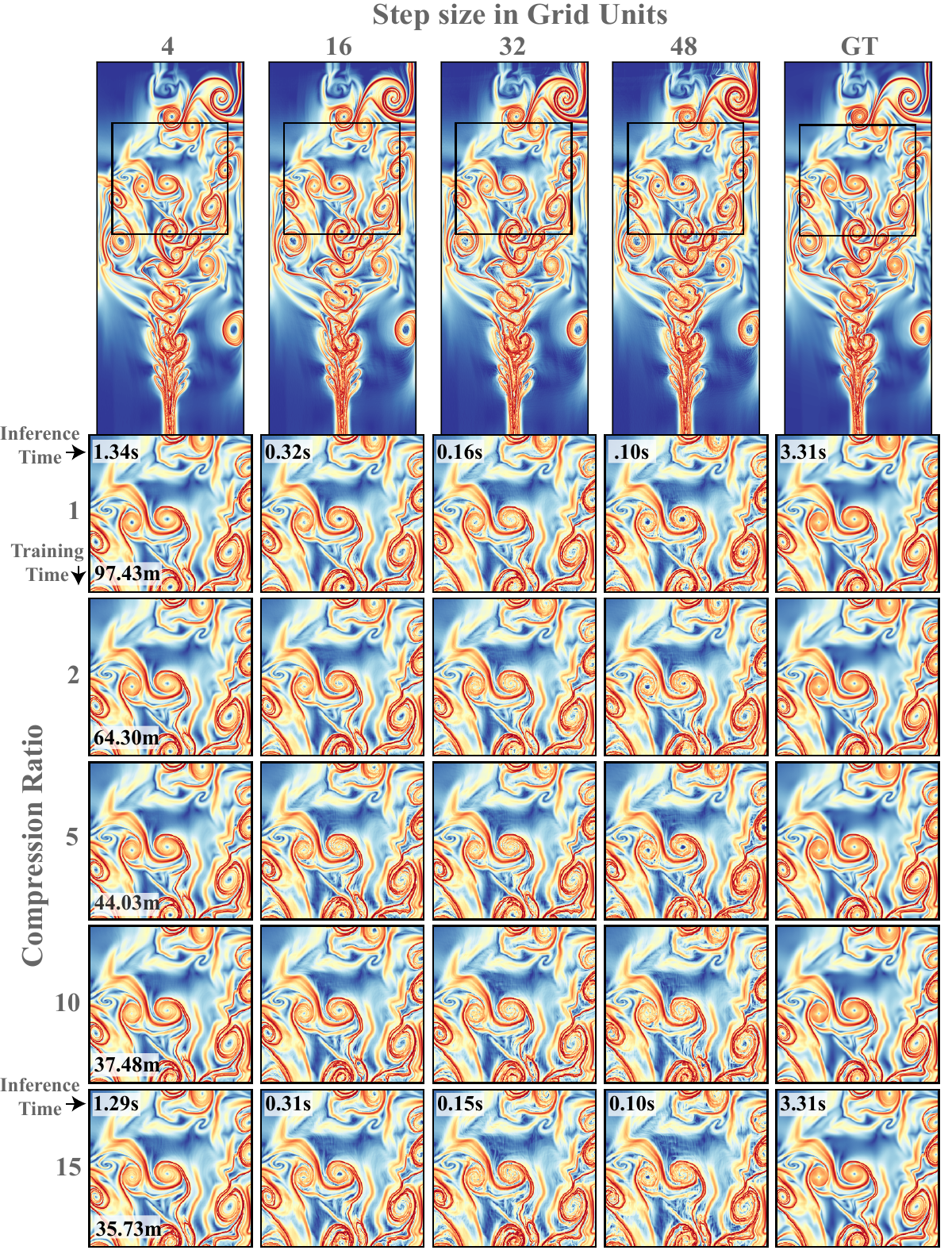}
    \caption{We qualitatively compare our model under varying compression ratios, showing the effect of compression on the step size taken by our model to produce the FTLE for the Boussinesq flow.} 
    \label{fig:grid_artifact}
\end{figure}

\begin{figure}[!t]
    \centering
    \includegraphics[width=1\linewidth]{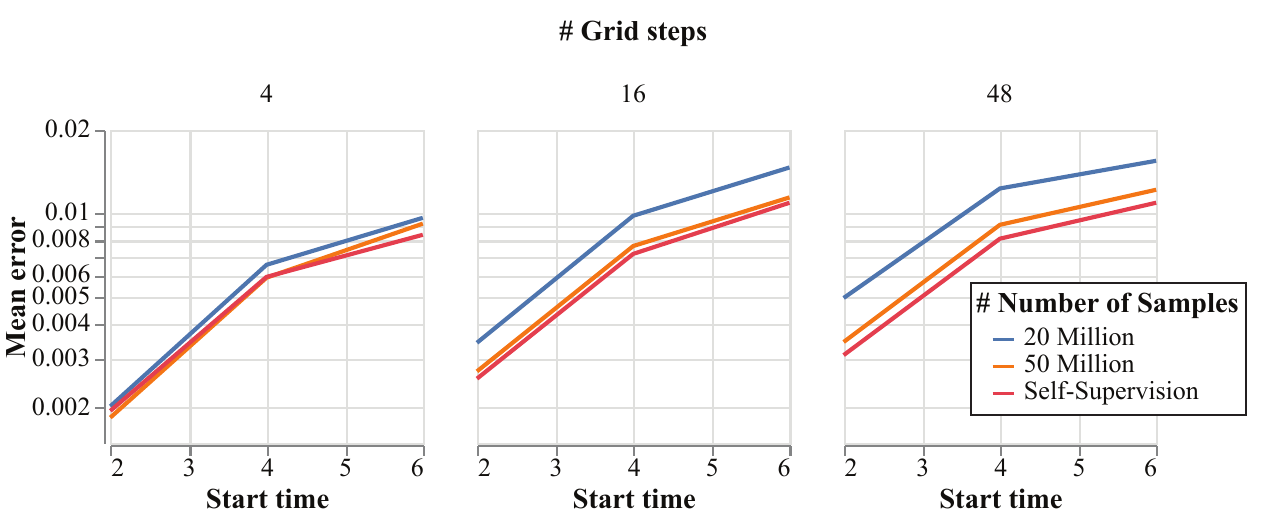}
    \caption{For the Boussinesq flow we compare our self-consistency criterion with that of directly supervising on flow samples, finding that our method produces comparable, if not improved, flow map approximations, without ever accessing the ground-truth flow map.} 
    \label{fig:self_consistency_exp}
\end{figure}

Our choice to learn flow maps via a self-supervisory signal is in contrast with how numerous visualization techniques interpolate~\cite{chandler2014interpolation,li2022efficient}, or build models~\cite{han2021exploratory} given samples of the flow map, e.g. typically as densely-sampled pathlines. Therefore we ask: is our self-consistency criterion an inferior objective to directly supervising on flow map samples? To this end, we have gathered a large collection of flow map samples, and modified our objective (Eq.~\ref{eq:composition-detail}) to accept the ground-truth flow map, and its corresponding derivative at the output position. We optimize for Boussinesq, using 20M and 50M flow map samples, and compare with our proposed objective, please see Fig.~\ref{fig:self_consistency_exp} for the results. We find that our method is able to learn comparable, if not better, flow map approximations, without ever observing flow map samples. In particular, at 50M samples we find that flow map supervision starts to become competitive with our method. Although supervising an on even larger number of samples might be more beneficial, clearly the data requirement starts to become prohibitively expensive, both for integrating the flow field, as well as storage requirements. In contrast, our method avoids these issues by requiring the vector field as the only supervision.

\section{Discussion}
In this paper we have presented an approach for integration free learning of flow maps, where we use coordinate-based neural networks as surrogates for fast and accurate flow map computation. We achieve this through our novel network design and optimization scheme that takes advantage of the basic properties of flow maps, in order to learn only using the provided vector field. We demonstrate the strength of our technique experimentally by comparing our method with various baselines and across multiple datasets.

There are several research directions we intend to pursue for future work. First, we acknowledge that although our method is scalable to optimize relative to other methods, optimization remains the computational bottleneck. We expect that porting our optimization scheme to the GPU, using fully-fused CUDA kernels for both the grid and MLPs, will alleviate this cost, as studied in prior works~\cite{weiss2021fast,muller2022instant}. Additionally, our self-consistency scheme is only an approximation, whereas other approaches have studied the design of invertible neural networks for computing discrete~\cite{behrmann2019invertible} or continuous~\cite{chen2018neuralode,bilovs2021neural} mappings of learned representations. We believe that adopting such approaches for representing flow maps in 2D and 3D unsteady flows is a fruitful research avenue.

% or


%\appendices

%\section{}

%\ifCLASSOPTIONcompsoc
%\else
%  % regular IEEE prefers the singular form
%\fi


\ifCLASSOPTIONcaptionsoff
  \newpage
\fi



\bibliographystyle{IEEEtran}
\bibliography{lagrangian, fast_ftle, inr, deep_learning_for_flow_vis, neural_ode}

\begin{thebibliography}{10}
\providecommand{\url}[1]{#1}
\csname url@samestyle\endcsname
\providecommand{\newblock}{\relax}
\providecommand{\bibinfo}[2]{#2}
\providecommand{\BIBentrySTDinterwordspacing}{\spaceskip=0pt\relax}
\providecommand{\BIBentryALTinterwordstretchfactor}{4}
\providecommand{\BIBentryALTinterwordspacing}{\spaceskip=\fontdimen2\font plus
\BIBentryALTinterwordstretchfactor\fontdimen3\font minus
  \fontdimen4\font\relax}
\providecommand{\BIBforeignlanguage}[2]{{%
\expandafter\ifx\csname l@#1\endcsname\relax
\typeout{** WARNING: IEEEtran.bst: No hyphenation pattern has been}%
\typeout{** loaded for the language `#1'. Using the pattern for}%
\typeout{** the default language instead.}%
\else
\language=\csname l@#1\endcsname
\fi
#2}}
\providecommand{\BIBdecl}{\relax}
\BIBdecl

\bibitem{haller2000finding}
G.~Haller, ``Finding finite-time invariant manifolds in two-dimensional
  velocity fields,'' \emph{Chaos: An Interdisciplinary Journal of Nonlinear
  Science}, vol.~10, no.~1, pp. 99--108, 2000.

\bibitem{haller2000lagrangian}
G.~Haller and G.~Yuan, ``Lagrangian coherent structures and mixing in
  two-dimensional turbulence,'' \emph{Physica D: Nonlinear Phenomena}, vol.
  147, no. 3-4, pp. 352--370, 2000.

\bibitem{cabral1993imaging}
B.~Cabral and L.~C. Leedom, ``Imaging vector fields using line integral
  convolution,'' in \emph{Proceedings of the 20th annual conference on Computer
  graphics and interactive techniques}, 1993, pp. 263--270.

\bibitem{froyland2009almost}
G.~Froyland and K.~Padberg, ``Almost-invariant sets and invariant
  manifolds—connecting probabilistic and geometric descriptions of coherent
  structures in flows,'' \emph{Physica D: Nonlinear Phenomena}, vol. 238,
  no.~16, pp. 1507--1523, 2009.

\bibitem{artale1997dispersion}
V.~Artale, G.~Boffetta, A.~Celani, M.~Cencini, and A.~Vulpiani, ``Dispersion of
  passive tracers in closed basins: Beyond the diffusion coefficient,''
  \emph{Physics of Fluids}, vol.~9, no.~11, pp. 3162--3171, 1997.

\bibitem{aurell1997predictability}
E.~Aurell, G.~Boffetta, A.~Crisanti, G.~Paladin, and A.~Vulpiani,
  ``Predictability in the large: an extension of the concept of lyapunov
  exponent,'' \emph{Journal of Physics A: Mathematical and General}, vol.~30,
  no.~1, p.~1, 1997.

\bibitem{you2014eulerian}
G.~You and S.~Leung, ``An eulerian method for computing the coherent ergodic
  partition of continuous dynamical systems,'' \emph{Journal of Computational
  Physics}, vol. 264, pp. 112--132, 2014.

\bibitem{garth2007efficient}
C.~Garth, F.~Gerhardt, X.~Tricoche, and H.~Hans, ``Efficient computation and
  visualization of coherent structures in fluid flow applications,'' \emph{IEEE
  Transactions on Visualization and Computer Graphics}, vol.~13, no.~6, pp.
  1464--1471, 2007.

\bibitem{kasten2009localized}
J.~Kasten, C.~Petz, I.~Hotz, B.~R. Noack, and H.-C. Hege, ``Localized
  finite-time lyapunov exponent for unsteady flow analysis.'' in \emph{VMV},
  2009, pp. 265--276.

\bibitem{sadlo2009visualizing}
F.~Sadlo and R.~Peikert, ``Visualizing lagrangian coherent structures and
  comparison to vector field topology,'' in \emph{Topology-Based Methods in
  Visualization II}.\hskip 1em plus 0.5em minus 0.4em\relax Springer, 2009, pp.
  15--29.

\bibitem{brunton2010fast}
S.~L. Brunton and C.~W. Rowley, ``Fast computation of finite-time lyapunov
  exponent fields for unsteady flows,'' \emph{Chaos: An Interdisciplinary
  Journal of Nonlinear Science}, vol.~20, no.~1, p. 017503, 2010.

\bibitem{lipinski2010ridge}
D.~Lipinski and K.~Mohseni, ``A ridge tracking algorithm and error estimate for
  efficient computation of lagrangian coherent structures,'' \emph{Chaos: An
  Interdisciplinary Journal of Nonlinear Science}, vol.~20, no.~1, p. 017504,
  2010.

\bibitem{sadlo2011time}
F.~Sadlo, A.~Rigazzi, and R.~Peikert, ``Time-dependent visualization of
  lagrangian coherent structures by grid advection,'' in \emph{Topological
  Methods in Data Analysis and Visualization}.\hskip 1em plus 0.5em minus
  0.4em\relax Springer, 2011, pp. 151--165.

\bibitem{stalling1995fast}
D.~Stalling and H.-C. Hege, ``Fast and resolution independent line integral
  convolution,'' in \emph{Proceedings of the 22nd annual conference on Computer
  graphics and interactive techniques}, 1995, pp. 249--256.

\bibitem{battke1997fast}
H.~Battke, D.~Stalling, and H.-C. Hege, ``Fast line integral convolution for
  arbitrary surfaces in 3d,'' in \emph{Visualization and Mathematics}.\hskip
  1em plus 0.5em minus 0.4em\relax Springer, 1997, pp. 181--195.

\bibitem{weinkauf2010streak}
T.~Weinkauf and H.~Theisel, ``Streak lines as tangent curves of a derived
  vector field,'' \emph{IEEE Transactions on Visualization and Computer
  Graphics}, vol.~16, no.~6, pp. 1225--1234, 2010.

\bibitem{han2021exploratory}
M.~Han, S.~Sane, and C.~R. Johnson, ``Exploratory lagrangian-based particle
  tracing using deep learning,'' \emph{arXiv preprint arXiv:2110.08338}, 2021.

\bibitem{bilovs2021neural}
M.~Bilo{\v{s}}, J.~Sommer, S.~S. Rangapuram, T.~Januschowski, and
  S.~G{\"u}nnemann, ``Neural flows: Efficient alternative to neural odes,''
  \emph{Advances in Neural Information Processing Systems}, vol.~34, pp.
  21\,325--21\,337, 2021.

\bibitem{muller2022instant}
T.~M{\"u}ller, A.~Evans, C.~Schied, and A.~Keller, ``Instant neural graphics
  primitives with a multiresolution hash encoding,'' \emph{arXiv preprint
  arXiv:2201.05989}, 2022.

\bibitem{agranovsky2014improved}
A.~Agranovsky, D.~Camp, C.~Garth, E.~W. Bethel, K.~I. Joy, and H.~Childs,
  ``Improved post hoc flow analysis via lagrangian representations,'' in
  \emph{2014 IEEE 4th Symposium on Large Data Analysis and Visualization
  (LDAV)}.\hskip 1em plus 0.5em minus 0.4em\relax IEEE, 2014, pp. 67--75.

\bibitem{chandler2014interpolation}
J.~Chandler, H.~Obermaier, and K.~I. Joy, ``Interpolation-based pathline
  tracing in particle-based flow visualization,'' \emph{IEEE transactions on
  visualization and computer graphics}, vol.~21, no.~1, pp. 68--80, 2014.

\bibitem{agranovsky2015multi}
A.~Agranovsky, H.~Obermaier, C.~Garth, and K.~I. Joy, ``A multi-resolution
  interpolation scheme for pathline based lagrangian flow representations,'' in
  \emph{Visualization and Data Analysis 2015}, vol. 9397.\hskip 1em plus 0.5em
  minus 0.4em\relax SPIE, 2015, pp. 221--232.

\bibitem{bujack2015lagrangian}
R.~Bujack and K.~I. Joy, ``Lagrangian representations of flow fields with
  parameter curves,'' in \emph{2015 IEEE 5th Symposium on Large Data Analysis
  and Visualization (LDAV)}.\hskip 1em plus 0.5em minus 0.4em\relax IEEE, 2015,
  pp. 41--48.

\bibitem{sane2019interpolation}
S.~Sane, H.~Childs, and R.~Bujack, ``An interpolation scheme for vdvp
  lagrangian basis flows.''

\bibitem{rapp2019void}
T.~Rapp, C.~Peters, and C.~Dachsbacher, ``Void-and-cluster sampling of large
  scattered data and trajectories,'' \emph{IEEE transactions on visualization
  and computer graphics}, vol.~26, no.~1, pp. 780--789, 2019.

\bibitem{chandler2016analysis}
J.~Chandler, R.~Bujack, and K.~I. Joy, ``Analysis of error in
  interpolation-based pathline tracing.'' in \emph{EuroVis (Short Papers)},
  2016, pp. 1--5.

\bibitem{hummel2016error}
M.~Hummel, R.~Bujack, K.~I. Joy, and C.~Garth, ``Error estimates for lagrangian
  flow field representations.'' in \emph{EuroVis (Short Papers)}, 2016, pp.
  7--11.

\bibitem{sane2018revisiting}
S.~Sane, R.~Bujack, and H.~Childs, ``Revisiting the evaluation of in situ
  lagrangian analysis.'' in \emph{EGPGV@ EuroVis}, 2018, pp. 63--67.

\bibitem{li2022efficient}
H.~Li, T.~Xiong, and H.-W. Shen, ``Efficient interpolation-based pathline
  tracing with b-spline curves in particle dataset,'' \emph{arXiv preprint
  arXiv:2207.07224}, 2022.

\bibitem{sadlo2007efficient}
F.~Sadlo and R.~Peikert, ``Efficient visualization of lagrangian coherent
  structures by filtered amr ridge extraction,'' \emph{IEEE Transactions on
  Visualization and Computer Graphics}, vol.~13, no.~6, pp. 1456--1463, 2007.

\bibitem{hlawatsch2010hierarchical}
M.~Hlawatsch, F.~Sadlo, and D.~Weiskopf, ``Hierarchical line integration,''
  \emph{IEEE transactions on visualization and computer graphics}, vol.~17,
  no.~8, pp. 1148--1163, 2010.

\bibitem{han2019flow}
J.~Han, J.~Tao, H.~Zheng, H.~Guo, D.~Z. Chen, and C.~Wang, ``Flow field
  reduction via reconstructing vector data from 3-d streamlines using deep
  learning,'' \emph{IEEE computer graphics and applications}, vol.~39, no.~4,
  pp. 54--67, 2019.

\bibitem{gu2021reconstructing}
P.~Gu, J.~Han, D.~Z. Chen, and C.~Wang, ``Reconstructing unsteady flow data
  from representative streamlines via diffusion and deep-learning-based
  denoising,'' \emph{IEEE Computer Graphics and Applications}, vol.~41, no.~6,
  pp. 111--121, 2021.

\bibitem{guo2020ssr}
L.~Guo, S.~Ye, J.~Han, H.~Zheng, H.~Gao, D.~Z. Chen, J.-X. Wang, and C.~Wang,
  ``Ssr-vfd: Spatial super-resolution for vector field data analysis and
  visualization,'' in \emph{2020 IEEE Pacific Visualization Symposium
  (PacificVis)}.\hskip 1em plus 0.5em minus 0.4em\relax IEEE Computer Society,
  2020, pp. 71--80.

\bibitem{sahoo2021integration}
S.~Sahoo and M.~Berger, ``Integration-aware vector field super resolution,''
  2021.

\bibitem{jakob2020fluid}
J.~Jakob, M.~Gross, and T.~G{\"u}nther, ``A fluid flow data set for machine
  learning and its application to neural flow map interpolation,'' \emph{IEEE
  Transactions on Visualization and Computer Graphics}, vol.~27, no.~2, pp.
  1279--1289, 2020.

\bibitem{chen2018neuralode}
R.~T.~Q. Chen, Y.~Rubanova, J.~Bettencourt, and D.~K. Duvenaud, ``Neural
  ordinary differential equations,'' in \emph{Advances in Neural Information
  Processing Systems}, vol.~31.\hskip 1em plus 0.5em minus 0.4em\relax Curran
  Associates, Inc., 2018.

\bibitem{liu2021second}
G.-H. Liu, T.~Chen, and E.~Theodorou, ``Second-order neural ode optimizer,''
  \emph{Advances in Neural Information Processing Systems}, vol.~34, pp.
  25\,267--25\,279, 2021.

\bibitem{norcliffe2021neural}
A.~Norcliffe, C.~Bodnar, B.~Day, J.~Moss, and P.~Li{\`o}, ``Neural ode
  processes,'' \emph{arXiv preprint arXiv:2103.12413}, 2021.

\bibitem{han2021temporal}
Z.~Han, Z.~Ding, Y.~Ma, Y.~Gu, and V.~Tresp, ``Temporal knowledge graph
  forecasting with neural ode,'' \emph{arXiv preprint arXiv:2101.05151}, 2021.

\bibitem{portwood2019turbulence}
G.~D. Portwood, P.~P. Mitra, M.~D. Ribeiro, T.~M. Nguyen, B.~T. Nadiga, J.~A.
  Saenz, M.~Chertkov, A.~Garg, A.~Anandkumar, A.~Dengel \emph{et~al.},
  ``Turbulence forecasting via neural ode,'' \emph{arXiv preprint
  arXiv:1911.05180}, 2019.

\bibitem{cai2020learning}
R.~Cai, G.~Yang, H.~Averbuch-Elor, Z.~Hao, S.~Belongie, N.~Snavely, and
  B.~Hariharan, ``Learning gradient fields for shape generation,'' in
  \emph{European Conference on Computer Vision}.\hskip 1em plus 0.5em minus
  0.4em\relax Springer, 2020, pp. 364--381.

\bibitem{lindell2021autoint}
D.~B. Lindell, J.~N. Martel, and G.~Wetzstein, ``Autoint: Automatic integration
  for fast neural volume rendering,'' in \emph{Proceedings of the IEEE/CVF
  Conference on Computer Vision and Pattern Recognition}, 2021, pp.
  14\,556--14\,565.

\bibitem{sahoo2022neural}
S.~Sahoo, Y.~Lu, and M.~Berger, ``Neural flow map reconstruction,'' in
  \emph{Computer Graphics Forum}, vol.~41, no.~3.\hskip 1em plus 0.5em minus
  0.4em\relax Wiley Online Library, 2022, pp. 391--402.

\bibitem{sitzmann2020implicit}
V.~Sitzmann, J.~Martel, A.~Bergman, D.~Lindell, and G.~Wetzstein, ``Implicit
  neural representations with periodic activation functions,'' \emph{Advances
  in Neural Information Processing Systems}, vol.~33, pp. 7462--7473, 2020.

\bibitem{tancik2020fourier}
M.~Tancik, P.~Srinivasan, B.~Mildenhall, S.~Fridovich-Keil, N.~Raghavan,
  U.~Singhal, R.~Ramamoorthi, J.~Barron, and R.~Ng, ``Fourier features let
  networks learn high frequency functions in low dimensional domains,''
  \emph{Advances in Neural Information Processing Systems}, vol.~33, pp.
  7537--7547, 2020.

\bibitem{fathony2021multiplicative}
R.~Fathony, A.~K. Sahu, D.~Willmott, and J.~Z. Kolter, ``Multiplicative filter
  networks,'' in \emph{International Conference on Learning Representations},
  2021.

\bibitem{takikawa2021neural}
T.~Takikawa, J.~Litalien, K.~Yin, K.~Kreis, C.~Loop, D.~Nowrouzezahrai,
  A.~Jacobson, M.~McGuire, and S.~Fidler, ``Neural geometric level of detail:
  Real-time rendering with implicit 3d shapes,'' in \emph{Proceedings of the
  IEEE/CVF Conference on Computer Vision and Pattern Recognition}, 2021, pp.
  11\,358--11\,367.

\bibitem{weiss2021fast}
S.~Weiss, P.~Herm{\"u}ller, and R.~Westermann, ``Fast neural representations
  for direct volume rendering,'' \emph{arXiv preprint arXiv:2112.01579}, 2021.

\bibitem{hayou2018selection}
S.~Hayou, A.~Doucet, and J.~Rousseau, ``On the selection of initialization and
  activation function for deep neural networks,'' \emph{arXiv preprint
  arXiv:1805.08266}, 2018.

\bibitem{kingma2015adam}
D.~P. Kingma and J.~Ba, ``Adam: A method for stochastic optimization,'' in
  \emph{ICLR (Poster)}, 2015.

\bibitem{behrmann2019invertible}
J.~Behrmann, W.~Grathwohl, R.~T. Chen, D.~Duvenaud, and J.-H. Jacobsen,
  ``Invertible residual networks,'' in \emph{International Conference on
  Machine Learning}.\hskip 1em plus 0.5em minus 0.4em\relax PMLR, 2019, pp.
  573--582.

\end{thebibliography}


% \begin{IEEEbiography}{Saroj Sahoo}
% \end{IEEEbiography}
\begin{IEEEbiography}{Saroj Sahoo}
Saroj Sahoo is currently a PhD student in the Computer Science department of Vanderbilt University. Prior to pursuing his PhD, he completed his B.Tech degree in Information Technology from Veer Surendra Sai University of Technology, India. His research interests are in leveraging deep learning techniques for scientific visualization.
\end{IEEEbiography}
\begin{IEEEbiography}{Matthew Berger}
Matthew Berger is currently an Assistant Professor in the Computer Science department of Vanderbilt University. He was previously a postdoc at the University of Arizona, received his PhD in Computing at the University of Utah in 2013, and received both his MS and BS in Computer Science from Binghamton University in 2007 and 2005, respectively. His research interests are in data visualization, visual analytics, and machine learning.
\end{IEEEbiography}

%\newpage



%\vfill


\appendices

\section{Flow Map Instantaneous Velocity}

In this appendix we show how our network design leads to a simple form for the flow map's instantaneous velocity.

First, we recall the specific form of our network. For clarity in derivations, we explicitly denote the dependency on time span $\tau$, and omit spatial position $\mathbf{x}$ and starting time $t$ where necessary. The first layer produces the following:
\begin{equation}
\label{eq:z0}
\mathbf{z}^{(0)}(\tau) = \sigma_{\nu}(\tau \mathbf{m}^{(0)}) \odot f_{\nu}(\mathbf{x},t).
\end{equation}
The function $f_{\nu} : \mathbb{R}^n \times \mathbb{R} \rightarrow \mathbb{R}^d$ is a $d$-dimensional spatiotemporal representation, for our purposes this is an arbitrary neural network. Vector $\mathbf{m}^{(0)} \in \mathbb{R}^d$ aims to reconcile domain-specific scaling, $\sigma_{\nu}$ is an activation function, and $\odot$ indicates element-wise multiplication. The second function $f_{\tau} : \mathbb{R}^n \times \mathbb{R} \rightarrow \mathbb{R}^d$ also learns a $d$-dimensional spatiotemporal representation. one specific to the flow map for nonzero time spans. These two representations are combined to give us the next layer's output:
\begin{align}
\label{eq:z1}
\mathbf{z}^{(1)}(\tau) & = \mathbf{z}^{(0)}(\tau) \\
& + \sigma_{\nu}\left(\tau \mathbf{m}^{(1)}\right) \odot \sigma_{\tau}\left(\mathbf{z}^{(0)}(\tau) \odot (W^{(1)} f_{\tau}(\mathbf{x},t)) \right) \nonumber ,
\end{align}
where $\mathbf{m}^{(1)} \in \mathbb{R}^d$ serves the same purpose as $\mathbf{m}^{(0)}$, and $W^{(1)} \in \mathbb{R}^{d \times d}$ is a learnable linear transformation. Subsequent representations are formed via residual connections:
\begin{equation}
\label{eq:zl}
\mathbf{z}^{(l)}(\tau) = \mathbf{z}^{(l-1)}(\tau) + \sigma_{\nu}\left(\tau \mathbf{m}^{(l)}\right) \odot \sigma_{\tau}\left(W^{(l)} \mathbf{z}^{(l-1)}(\tau) \right),
\end{equation}
while the last layer $L$ applies a single linear transformation to give us the output position, wherein we also include a skip connection for the input position:
\begin{equation}
\label{eq:zlast}
\hat{\Phi}(\mathbf{x},t,\tau) = \mathbf{x} + W^{(L)} \mathbf{z}^{(L-1)}(\tau).
\end{equation}
We further make the following assumptions on activation functions $\sigma_{\nu}$ and $\sigma_{\tau}$:
\begin{align}
\sigma_{\nu}(0) = 0 \nonumber \\
\sigma'_{\nu}(0) \ne 0 \nonumber \\
\sigma_{\tau}(0) = 0 \nonumber
\end{align}

We aim to compute the derivative of the neural flow map at time span $\tau = 0$:
\begin{equation}
\label{eq:total}
\frac{d \hat{\Phi}(\mathbf{x},t,0)}{d \tau} = \frac{d \hat{\Phi}}{d \mathbf{z}^{(L-1)}} \frac{d \mathbf{z}^{(L-1)}(0)}{d \tau}.
\end{equation}
The term on the left is simply:
\begin{equation}
\label{eq:simpledphi}
\frac{d \hat{\Phi}}{d \mathbf{z}^{(L-1)}}  = W^{(L)}
\end{equation}
As for the term on the right, we have the following recurrence for $l > 1$:
\begin{align}
\frac{d \mathbf{z}^{(l)}}{d \tau} & = \frac{d \mathbf{z}^{(l-1)}(0)}{d \tau} \\
& + \sigma_v'(0) \mathbf{m}^{(l)} \odot \sigma_{\tau}\left(W^{(l)} \mathbf{z}^{(l-1)} \nonumber  \right)\\
& + \sigma_{\nu}(0) \odot \frac{d \sigma_{\tau}\left(W^{(l)} \mathbf{z}^{(l-1)} \right)}{d \tau} \nonumber .
\end{align}
The term in the second line will evaluate to zero, since activation vectors $\mathbf{z}^{(l)}(0)$, for any layer $l$, will be zero due to the multiplicative scaling with $\tau = 0$, and the fact that $\sigma_{\tau}(0) = 0$. Likewise, the third line will vanish due to our assumption on the activation function evaluating to $\sigma_{\nu}(0) = 0$. A similar reasoning can be applied for layer $l = 1$ (c.f. Eq.~\ref{eq:z1}), due to the multiplicative scaling of $\mathbf{z}^{(0)}(0)$ with $(W^{(1)} f_{\tau}(\mathbf{x},t))$. Thus, we have the following:
\begin{equation}
\label{eq:simpledz}
\frac{d \mathbf{z}^{(L-1)}(0)}{d \tau} = \frac{d \mathbf{z}^{(0)}(0)}{d \tau} = \left(\sigma_{\nu}'(0) \mathbf{m}^{(0)}\right) \odot f_{\nu}(\mathbf{x},t).
\end{equation}

As we choose $\sigma_{\nu}$ to be a hyperbolic tangent function, we obtain $\sigma_{\nu}'(0) = 1$. Thus, plugging Eqs.~\ref{eq:simpledphi} and~\ref{eq:simpledz} into Eq.~\ref{eq:total}, we arrive at:
\begin{equation}
\frac{d \hat{\Phi}(\mathbf{x},t,0)}{d \tau} = W^{(L)} (\mathbf{m}^{(0)} \odot f_{\nu}(\mathbf{x},t)).
\end{equation}
Note that in Eq.~\ref{eq:z1}, the multiplicative scaling within the activation $\sigma_{\tau}$ is essential for this result - a different way of combining the representations, e.g. adding them together, would introduce dependencies on $f_{\tau}$, and weight matrices $W^{(l)}$, $l > 0$, in computing the derivative.

\begin{figure}[!t]
\centering
\includegraphics[width=1\linewidth]{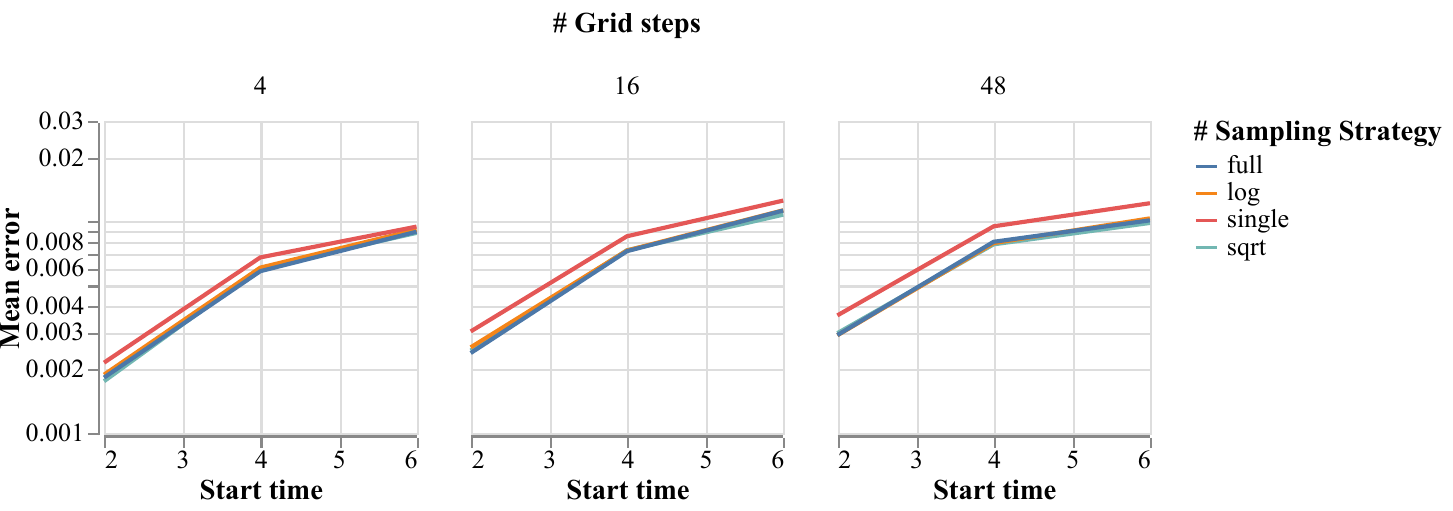}
\caption{We study the effect of different schemes for taking steps in flow map optimization for the Boussinesq flow: full corresponds to taking one step per grid unit in time (and thus the most costly), sqrt corresponds to a square root scaling in (grid unit) time, log is a logarithmic scaling, while single corresponds to taking just a single step (thus the cheapest in computation). We find little difference between these schemes upon evaluation.} 
\label{fig:sampleboussinesq}
\end{figure}

\begin{figure}[!t]
\centering
\includegraphics[width=1\linewidth]{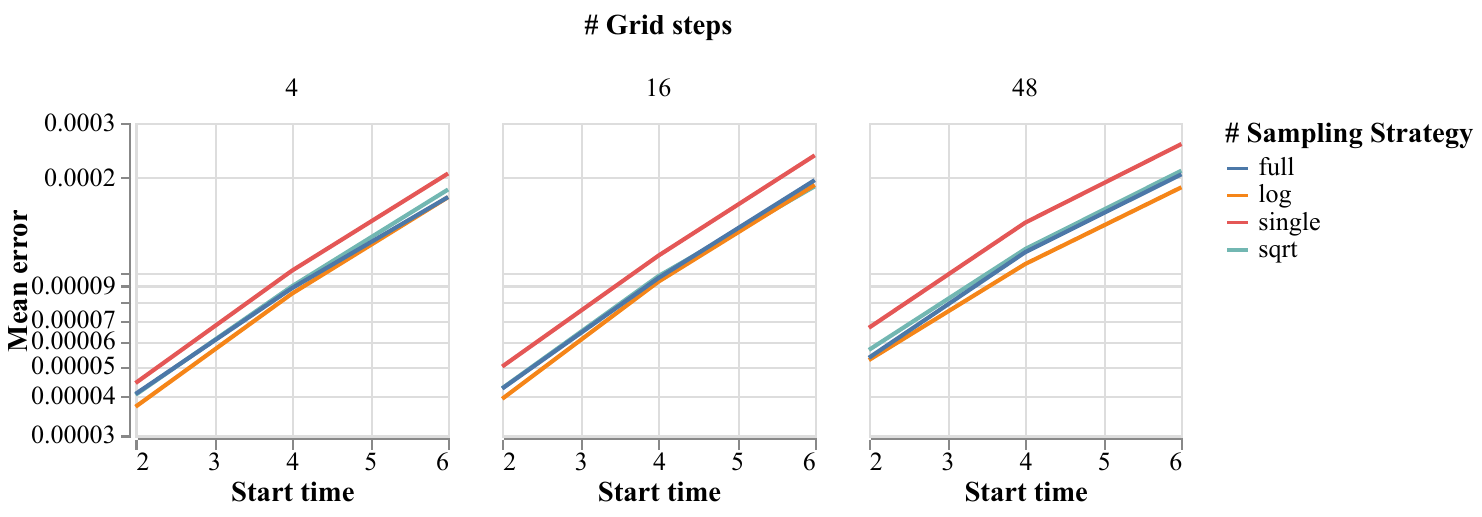}
\caption{We further study different schemes for taking steps in flow map optimization, here for Double Gyre.}
\label{fig:sampledoublegyre}
\end{figure}

\begin{figure}[!t]
\centering
\includegraphics[width=1\linewidth]{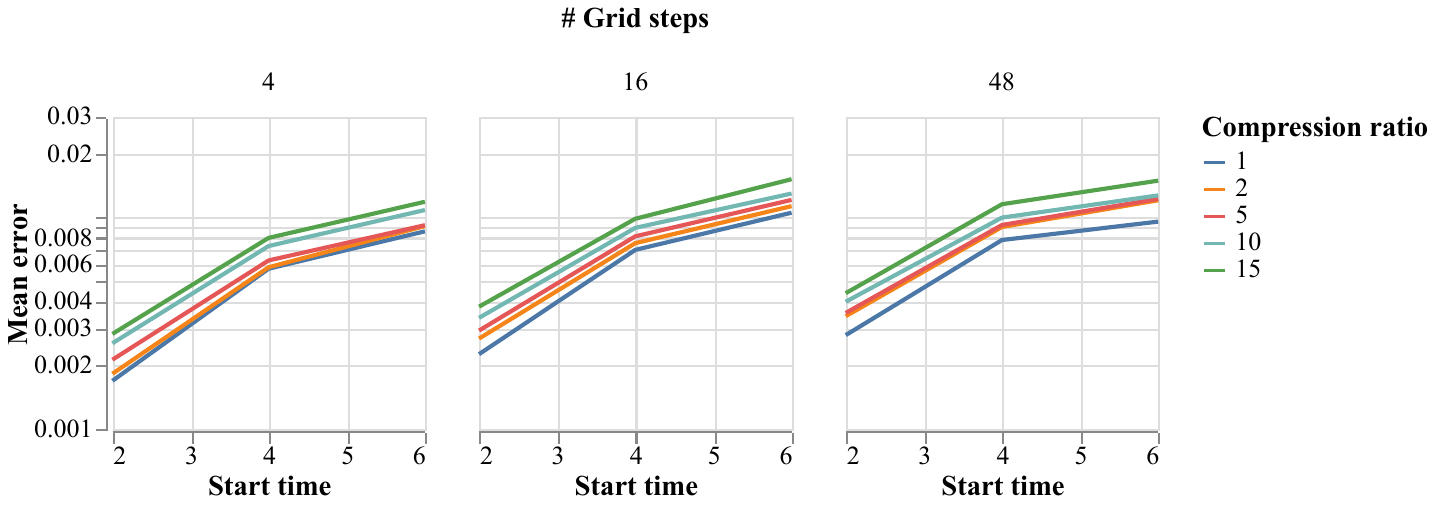}
\caption{We study the effect of model size on accuracy for the Boussinesq flow. We find a small drop in accuracy as we decrease the model size.} 
\label{fig:compression}
\end{figure}

\begin{figure}[!t]
\centering
\includegraphics[width=1\linewidth]{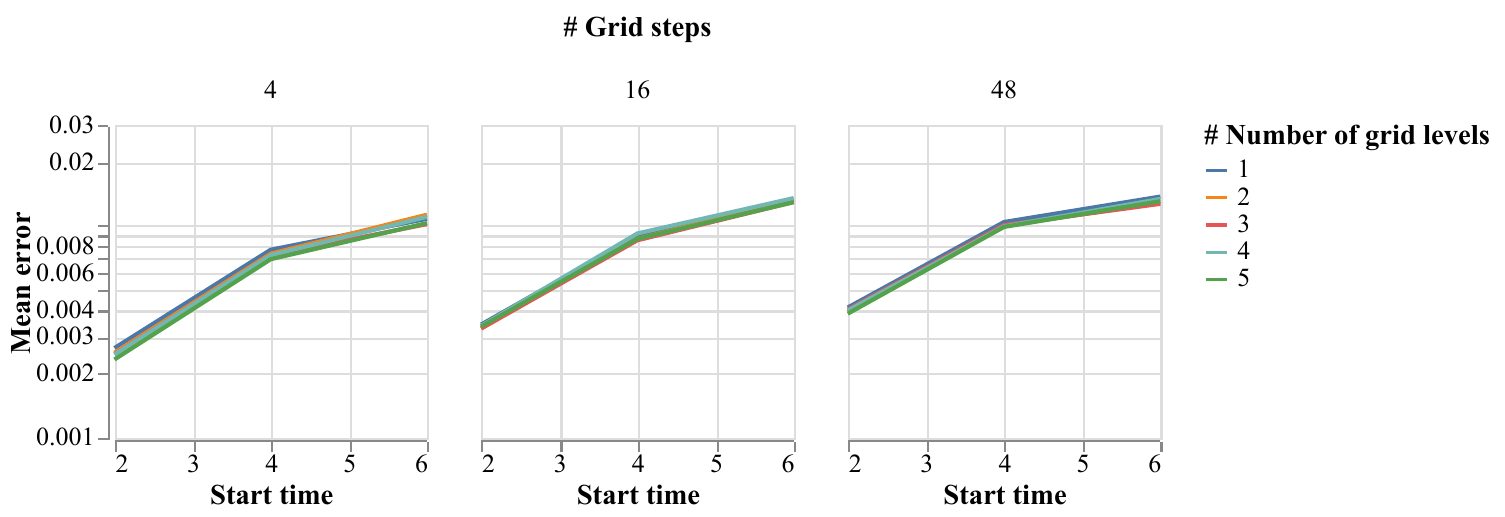}
\caption{We vary the number of grid levels used in the multiresolution feature grid for Boussinesq, and find that our method is robust to this particular hyperparameter setting.} 
\label{fig:levels}
\end{figure}

\begin{figure}[!t]
    \centering
    \includegraphics[width=.4\linewidth]{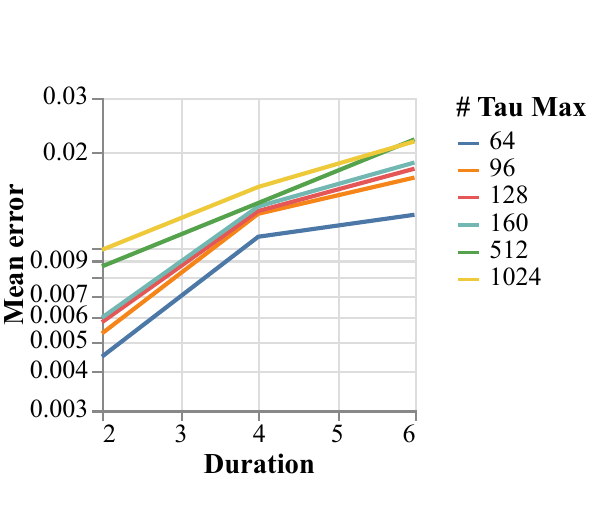}
    \caption{For the Boussinesq flow we compare the effects of $\tau_{max}$ on the overall performance of the model, finding that the self-consistency criterion when trained with large $\tau_{max}$ gracefully degrades in performance, indicating the stability of our optimization technique.} 
    \label{fig:tau_max}
\end{figure}
\section{Ablation results}
We include model/optimization ablation results to demonstrate the robustness of our method across a variety of parameter settings. In all experiments the maximum time span $\tau_{max}$ is set to 48 (in grid units), and we use default parameters as originally specified in the paper, unless otherwise mentioned.

Figs.~\ref{fig:sampleboussinesq} and~\ref{fig:sampledoublegyre} compare different strategies for the number of steps taken by our method in optimizing the self-consistency criterion: one step per grid unit, a square root scaling (the default choice used throughout all results in the paper), a log scaling, as well as taking just a single step. As shown in the figures, across datasets we find little difference in the results, evaluated across varying step sizes. As a compromise, we set the square root scaling as it adds little computation cost, while ensuring additional stability in optimization.

\begin{figure}[!t]
    \centering
    \includegraphics[width=1\linewidth]{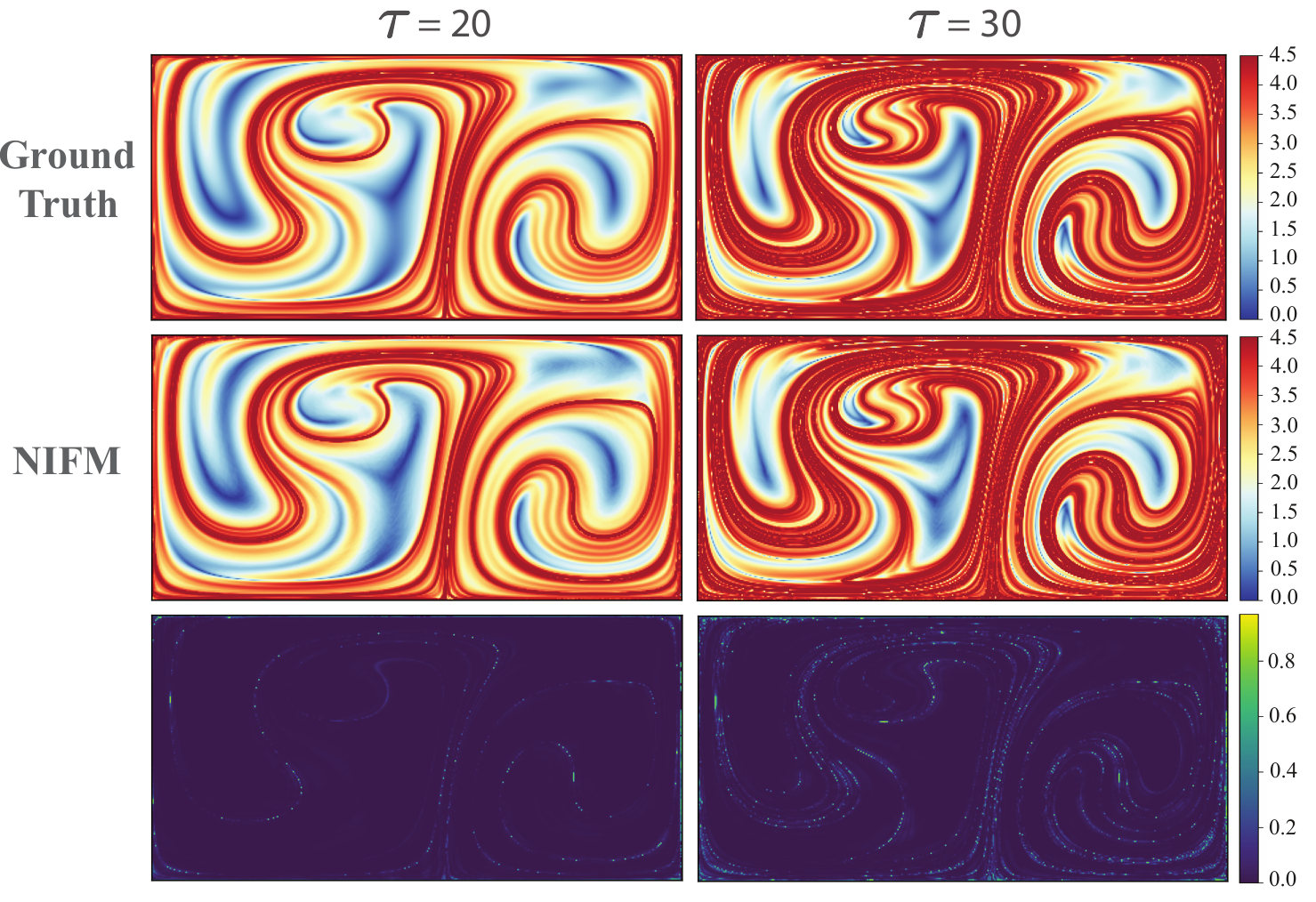}
    \caption{We show the performance of NIFM under large integration durations for the double gyre dataset. We show the FTLE and the corresponding flow map errors for $\tau=20$ and $\tau=30$. We observe the technique is able to perform reasonably well even for very large integration durations.} 
    \label{fig:dg_5x}
\end{figure}

In Fig.~\ref{fig:compression} we study the effect of model size, e.g. compression ratio, on accuracy for the Boussinesq flow. In general we find a small drop in accuracy, suggesting that our model can generalize well even when utilizing a smaller number of parameters. In Fig.~\ref{fig:levels} we study the impact on the number of grid levels used for our multiresolution feature grid. Although we default the number of levels to $4$ in the main paper, in general, we find little difference in quality as we adjust the number of levels. In Fig.~\ref{fig:tau_max} we study the impact of $\tau_{max}$ on the performance of the model. We evaluate the model by taking the minimum between the integration duration and the $\tau_{max}$ the model was trained on as the step size. We found that when trained with large values of $\tau_{max}$ the overall performance of the model degrades affecting both smaller and larger timespans. Last, in Fig.~\ref{fig:dg_5x} we study the performance of NIFM when the integration duration is large. We perform this experiment on the double gyre dataset and evaluate the model for $\tau=20$ and $\tau=30$  respectively. From the FTLE and its corresponding flow map errors, we can see that even for long integration duration the model performs reasonably well.

\end{document}